%% file: Ver1.tex
\pdfoutput=1
\documentclass[acmsmall]{acmart}

\newcommand{\ra}{\rightarrow} 

\newcommand{\da}{\downarrow}

\newcommand{\Exp}[1]{\mathbb{E}\left[#1\right]} 

\newcommand{\ignore}[1]{}
\def\opt{\mathsf{OPT}}

\usepackage{enumitem,kantlipsum}

\usepackage{tikz}
\usepackage{pgfplots}
\definecolor{bblue}{HTML}{4F81BD}
\definecolor{rred}{HTML}{C0504D}
\definecolor{ggreen}{HTML}{9BBB59}
\definecolor{ppurple}{HTML}{9F4C7C}
\usepackage{comment}
\usepackage{nicefrac}
\usepackage{ifthen}

\usetikzlibrary{calc}

\input{input.tex}

\newboolean{showcomments}
\setboolean{showcomments}{true}
\newcommand{\jk}[1]{  \ifthenelse{\boolean{showcomments}}
{ \textcolor{red}{(JK says:  #1)}} {}  }
\newcommand{\rahul}[1]{  \ifthenelse{\boolean{showcomments}}
{ \textcolor{green}{(RV says:  #1)}} {}  }

\newlength{\figurewidth}

\newcounter{one}
\newcounter{two}
\usepackage{amsmath}
\usepackage{amsfonts}
\begin{document}
\title{Capacity Provisioning Motivated Online Non-Convex
  Optimization Problem with Memory and Switching Cost}
\author{Rahul Vaze, Jayakrishnan Nair}
\begin{abstract}
An online non-convex optimization problem  is considered where the
goal is to minimize the flow time (total delay) of a set of jobs by
modulating the number of active servers, but with a switching cost
associated with changing the number of active servers over time. Each
job can be processed by at most one fixed speed server at any
time. Compared to the usual online convex optimization (OCO) problem with switching cost, the
objective function considered is non-convex and more importantly, at each time, it depends on all past
decisions and not just the present one. Both worst-case and stochastic
inputs are considered; for both cases, competitive algorithms
are derived.
\end{abstract}
\maketitle


\input{Intro}

\input{ProblemFormulation}


\input{stochastic}
\input{burst_opt}
\input{SimResultsToBeAdded}

\input{Conclusions}
\bibliographystyle{elsarticle-num} 
\bibliography{refs,references,SpeedScaling}

\input{MissingProofs}
\input{CorrectDualLowerBound}
\input{AppLemmaFlowTime}

\end{document}

%% file: input.tex
\usepackage{amsfonts}
\newtheorem{theorem}{Theorem}

\newtheorem{definition}[theorem]{Definition}

\newtheorem{lemma}[theorem]{Lemma}

\newtheorem{proposition}[theorem]{Proposition}

\newtheorem{remark}[theorem]{Remark}

\def\bb0{{\mathbb{0}}}

\def\bb{{\mathbf{b}}}

\def\b0{{\mathbf{0}}}
\def\alg{\mathsf{ALG}}
\def\opt{\mathsf{OPT}}

\def\bP{{\mathbf{P}}}

\def\b1{{\mathbf{1}}}

\def\bbR{{\mathbb{R}}}

\def\bbZ{{\mathbb{Z}}}

\def\cA{\mathcal{A}}
\def\cB{\mathcal{B}}

\def\cI{\mathcal{I}}
\def\cJ{\mathcal{J}}

\def\cL{\mathcal{L}}

\def\cO{\mathcal{O}}

\def\sfA{\mathsf{A}}

\def\sfC{\mathsf{C}}

\def\sfN{\mathsf{N}}

\def\sfd{{\mathsf{d}}}

\def\sf0{{\mathsf{0}}}

\def\nn{\nonumber}

%% file: Intro.tex
\section{Introduction}

The capacity provisioning problem introduced in \cite{lin2012online} models a canonical resource allocation problem in data centres, that is described as follows. Let the demand at time $t$ be $d(t)$ that
is revealed causally over time.  Dedicating $s(t)$ number of servers
at time $t$ to support this demand, two types of costs are incurred;
the quality of service cost depending on $d(t)$ and $s(t),$ and the
energy cost required to run the $s(t)$ servers. The two costs are
combined to produce a single cost function $f_t(s(t), d(t))$. In
addition, because of obvious practical limitations, there is a penalty
associated with changing the number of servers across time slots,
captured by cost function $c(s(t-1), s(t))$, where $c$ is a specific
function that increases with $|s(t-1)-s(t)|$. Overall, the
optimization problem is to choose $s(t)$ with uncertain demand $d(t)$
so as to minimize $\sum_t f_t(s(t), d(t)) + c(s(t-1), s(t))$. 

To study this powerful and important formulation, a problem called OCO-S was formulated and studied extensively, where at each discrete time $t$, a
convex cost function $f_t: \bbR^{\sfd} \rightarrow \bbR^+$ is revealed, and an
algorithm has to choose an action~$x_t$ knowing all $f_\tau, \tau \le
t$. The cost of choosing action $x_t$ at time $t$ is the sum of the cost $f_t(x_t)$, and the switching cost
$c(x_{t-1}, x_t)$. The goal is to choose $x_t, 1\le t\le T$ such that
the overall cost 
\begin{equation}\label{eq:oco}
\sum_{t=1}^T f_t(x_t) + c(x_{t-1}, x_t)
\vspace{-0.08in}
\end{equation}
 is minimized.  
The capacity provisioning problem is a special case of \eqref{eq:oco} with $\sfd = 1$. More notably, since $f_t$'s are convex, $x_t$'s are required to be real numbers, OCO-S models the capacity provisioning problem only when $s(t)$ and $d(t)$ are not enforced to be integers. We will overcome this limitation in this paper for a specific cost function.

Even though the capacity provisioning problem \cite{lin2012online} is a powerful abstraction, however, it absorbed all
the QoS performance metrics, e.g. delay/latency within a single
slot. In particular, QoS part of the objective function $f_t(s(t),
d(t))$ did not depend on job arrivals and decisions made in previous time
slots, and only the switching cost accounted for changing decisions
across time slots. It essentially took an averaged-out view that a
time slot is long enough and restricted all the queueing dynamics to
happen within that slot. 

For any practical system, however, the cost incurred at any time also depends 
on some or all past decisions, and to model this important memory feature, we study a data center relevant QoS centric problem defined as follows. 
We consider a canonical problem for data centers where jobs arrive over time and the QoS metric of interest is the flow-time (sum of the delay seen by each job). Each server has a fixed speed and can serve at most one job at a time.
 Similar to the OCO-S, there is a switching cost in changing
 the number of active servers across time slots, and the goal is to minimize the flow time plus the
 overall switching cost, where the decision variable at each time is the 
 number of active servers.

The flow time $F = \sum_t n(t)$, where $n(t)$ is the number of outstanding jobs  at time $t$. Thus, formally the problem is 
 \begin{equation}\label{eq:introocoflowtime1} \min_{s(t)}  \sum_{t} n(t) + \alpha \sum_{t}  c(s(t), s(t-1)),
\end{equation}
where $n(t) =
\max\left\{\sum_{\tau\leq t} \text{arr}_\tau - \sum_{\tau \leq t-1}
\text{dep}_\tau,0\right\}$, and $\text{arr}_\tau$ and $\text{dep}_\tau$ are the
number of jobs that arrive at the start of slot $\tau$ and depart at
the end of slot $\tau$, respectively, and $s(t)$ is the number of active servers at time $t$, and $\alpha>0$ is the weighting parameter. 

Since $\text{dep}_\tau$ depends
on the decisions about all the number of active servers used so far, the
objective function at time $t$, $n(t)$, depends on all the decisions
made so far. Moreover, the objective function at time $t$, i.e.,
$n(t)$ (to model the flow time), {\bf is not convex since the number of servers and the number of outstanding jobs at any time are both integers}. 

The goal of an online algorithm \(\cA\) is to choose $s(t)$ to minimize the competitive
ratio defined as 
$$\text{cr}_\cA = \max_{\sigma}\frac{C_\cA(\sigma)}{C_\opt(\sigma)},$$
where $\sigma$ is the input, $C_\cA$ \eqref{eq:introocoflowtime1} is the cost of $\cA$, while \(C_\opt\) denotes the cost
of the optimal offline algorithm $\opt$, that knows the input $\sigma$
ahead of time, non-causally.

%

\subsection{Prior Work}
\subsubsection{OCO without memory}
Online convex optimization (OCO) is a standard framework for modelling and analyzing a broad family of online decision problems under uncertainty. With OCO, at time $t$ an online policy selects an action $x_t$ first, and then the adversary reveals a convex cost function $f_t$, and the cost of $f_t(x_t)$ is incurred. 
The performance metric is static or dynamic regret, depending on whether the benchmark is allowed to choose a single action or can
one action for each time slot $t$. In particular, 
\vspace{-0.1in}
\begin{eqnarray} \label{def:statregret-def}
	\textrm{Static Regret}_T \equiv \sup_{\{f_t\}_{t=1}^T} \sup_{x^\star \in \mathcal{X}}  \sum_{t=1}^T f_t(x_t) - \sum_{t=1}^T f_t(x^\star).
\end{eqnarray}
 The static-ness refers to the benchmark using only one action $x^\star$ throughout the horizon.
\begin{eqnarray} \label{def:statregret-def}
	\textrm{Dynamic Regret}_T \equiv \sup_{\{f_t\}_{t=1}^T} \sup_{x_t^\star \in \mathcal{X}} \sum_{t=1}^T f_t(x_t) - \sum_{t=1}^T f_t(x_t^\star).
\end{eqnarray}
There is a large body of work on OCO for both static and dynamic regret, and we refer the reader to \cite{orabona2019modern} for a textbook treatment. 

\subsubsection{OCO with memory}
OCO framework has been extended to include cost functions with memory, i.e., $f_t$'s do not just depend on $x_t$ but also on past decisions. In particular, \cite{anava2013online} considered $p$-length memory cost functions where $f_t(x^t) = f_t(x_t, x_{t-1}, \dots,x_{t-p+1})$, i.e. the current cost 
 function depends on the last $p$ actions. It was shown in \cite{anava2013online}, that the FTRL algorithm has a static regret of $O(p^{3/2} \sqrt{T})$, which has been improved in \cite{kumar2023online} to $O(p \sqrt{T})$, which has also been shown to be tight, i.e. no algorithm can get a better static regret than 
 $\Omega(p \sqrt{T})$. In particular, \cite{kumar2023online} provides bad news that the regret grows linearly in the memory length. This also implies that the competitive ratio of any algorithm for solving OCO-S \eqref{eq:oco} should grow linearly with $p$ for arbitrary convex functions with memory.

\subsubsection{OCO-S}

OCO-S \eqref{eq:oco} is a  generalization of OCO, however, with OCO-S, $x_t$ is allowed to be chosen after $f_t$ is revealed, since the presence of switching cost keeps the problem non-trivial. 
For the OCO-S \eqref{eq:oco} with quadratic switching cost, i.e., $c(x_{t},
x_{t - 1}) = ||x_{t} - x_{t - 1}|| ^ 2$, if the functions \(f_{t}\)'s
are just convex, then the competitive ratio of any
deterministic online algorithm is unbounded
\cite{goel2019beyond}. Thus, starting with \cite{chen2018}, to solve
\eqref{eq:oco}, $f_t$'s were assumed to be strongly convex. With
\(\mu\)-strongly convex $f_t$'s, online balanced descent (OBD)
algorithm that balances the suboptimality and switching cost at each
time, has a competitive ratio of at most \(3 +
\mathcal{O}(\frac{1}{\mu})\) \cite{goel2019online}. Surprisingly, \cite{ZhangSOCO2021XstarAlgorithm} showed that the same guarantee can be 
obtained by a simple algorithm that chooses $x_t$ as the local optimizer of $f_t$ disregarding the switching cost. This result was
improved in \cite{goel2019beyond} using a modification of OBD to get a
competitive ratio $\mathcal{O}(\frac{1}{\sqrt{\mu}})$ as \({\mu \to
  0^{+}}\), which is also shown to be the best possible (order-wise)
competitive ratio for all possible deterministic online algorithms
\cite{goel2019beyond}. 
Similar results have been obtained 
when $f_t$'s are locally polyhedral in \cite{chen2018}. 
Very recently, \cite{bhuyan2023best} studied OCO-S under some assumptions on the sequence of minimizers of $f_t$'s and provided an algorithm that works for both the worst case as well as the stochastic input. The OCO-S has also been studied in the information setting of OCO recently \cite{senapativaze2023}, where action $x_t$ is chosen before $f_t$ is revealed.

OCO-S with linear switching cost, where $c(x_{t}, x_{t - 1})= |x_{t}
- x_{t - 1}|$ has also been widely considered. For this case, an
optimal offline algorithm was derived in \cite{lin2012online}, while
online algorithms have been explored in \cite{lin2012online,
  lin2012dynamic, andrew2013tale, bansal20152,
  DBLP:journals/corr/abs-1807-05112,chen2018, argue2020dimension, chen2018smoothed} when
$f_t$'s are just convex.  For $\sfd=1$, the best known online algorithm
has the competitive ratio of $2$ which matches the lower bound
\cite{DBLP:journals/corr/abs-1807-05112}. It is important to notice
that unlike the quadratic switching cost case, when the switching cost
is linear, the $f_t$'s are not required to be strongly convex to
get meaningful competitive ratios.  For general $\sfd$, when \(f_{t}\)'s
are convex, the competitive ratio of any algorithm has a lower bound
\(\Omega(\sqrt{\sfd})\) \cite{chen2018}. Imposing that \(f_{t}\)'s are
\(\mu\)-strongly convex and \(L\)-smooth, defining \(\kappa := L /
\mu\) as the condition number, \cite{argue2020dimension} proposed a
constrained online balanced descent (COBD) algorithm with a competitive
ratio \(\cO(\sqrt{\kappa})\). Recently, \cite{soco-constraints} considered the OCO-S with linear switching cost under some long-term constraints.
OCO-S with both trusted and untrusted predictions has also been considered in \cite{chen2015online, chen2016using, li2018using, christianson2022chasing, rutten2023smoothed}.
A review on OCO-S can be found in textbook \cite{vazebook}.

OCO-S with memory has also been studied \cite{shi2020online} where memory is introduced only in the switching cost, i.e.  $f_t( x^t)=f_t(x_t)$ but switching cost $c(s_t,s_{t-1}) = ||s_t - \sum_{k=1}^p \gamma_k s_{t-k}||^2$, i.e. it depends on the last $p$ actions when $f_t$ are strongly convex. In comparison, problem \eqref{eq:introocoflowtime1} has memory in the cost functions and cost functions are non-convex.
We would like to note here that OCO-S is a very well-studied problem and the reference list provided is not exhaustive.

\subsubsection{Speed scaling}
Instead of counting switching cost with fixed speed servers like in \eqref{eq:introocoflowtime1}, there is an alternate well-studied formulation called 
speed-scaling when the server speeds are tuneable. Operating a server at speed $s$ incurs a cost of $P(s)$, where $P$ is an increasing convex 
increasing. The objective is to minimize the flow time + total energy spent. 
 The flow-time plus energy problem has been studied in \cite{andrew2010optimality, surveyspeedscaling, bansal2008scheduling, bansal, bansal2010speed, SpeedScalingPS, SpeedScalingPEVA}. 
Recently, \cite{rutten2021new} considered the machine learning augmented setup when the job arrival sequence is stochastic and proposed a consistent, smooth, and robust algorithm.

The multi-server case with homogenous servers (all servers have the same power function) has been studied in \cite{vaze2020multiple} while  \cite{multiserverspeedscaling, devanur2017primal} study the heterogeneous case. 
The most general system model with speed-scaling has been 
studied in \cite{vazenetwork}, where there is a network of speed tuneable servers, and the objective is to minimize the sum of flow time plus energy for  multiple source-destination pairs. 
The speed-scaling model is fundamentally different than the considered problem \eqref{eq:introocoflowtime1}, where $s(t)$ is discrete and the switching cost is counted incrementally, i.e. it depends on $|s(t)-s(t-1)|$.

\vspace{-0.15in}
\subsection{Comparison with Prior Work}

At this point it is instructive to recall the lower bound of $\Omega(p \sqrt{T})$ from \cite{kumar2023online} on the static regret of any online algorithm for OCO with length $p$ memory. Given that the problem \eqref{eq:introocoflowtime1} that we want to study has unbounded $p$ and is non-convex, and the benchmark ($\opt$) is allowed to choose distinct actions for each time slot, it might appear that the performance of any algorithm for solving \eqref{eq:introocoflowtime1} will be arbitrarily bad. By carefully exploiting the structure of the problem \eqref{eq:introocoflowtime1}, we will avoid the obvious bad results one would expect with unbounded memory and non-convex cost functions. We will show memory length independent results for problem \eqref{eq:introocoflowtime1}.


Conceptually,  the principal difference between the problem we consider in this paper
 and the capacity provisioning problem \cite{lin2012online} is the
 time scale under consideration. The capacity provisioning problem
 takes an averaged-out view that a time slot is long enough and all
 the queuing dynamics are assumed to play out within it, while we are
 taking a zoomed in view and are interested in modelling the queuing
 dynamics explicitly and study its interaction with the switching
 cost. 
 From an applications point of view, our approach is valid for
 short time scale settings where dynamics evolve at a rapid rate. Analytically, the difficulty is because of the introduction of memory in the cost functions across time as well as the non-convexity of the 
 cost function.

\subsection{Our Contributions}
In this paper, we consider both the worst-case and stochastic models
for job arrivals and the objective is to design algorithms for solving \eqref{eq:introocoflowtime1} with small
competitive ratios. 
Our contributions are as follows. At time~$t$, let $n(t)$ and $s(t)$ denote the number
of outstanding jobs and the number of active servers, respectively.
\begin{enumerate}[wide, labelwidth=!, labelindent=0pt]
\item Worst case input: In this case, we assume that all job sizes are identical, which is still a highly challenging problem. 
\begin{itemize}
\item When the switching cost is linear and $\alpha\le 1$, we show that a simple
  algorithm that chooses the number of active servers equal to the
  number of outstanding jobs ($s(t) = n(t)$) achieves a competitive
  ratio of $2$. We also construct an input instance where this competitive ratio
  guarantee is tight for this algorithm.
  \item When the switching cost is linear and $\alpha>1$, we show that natural extensions of algorithms  \cite{lin2012online,
  lin2012dynamic, andrew2013tale, bansal20152,
  DBLP:journals/corr/abs-1807-05112,chen2018, argue2020dimension}
  known to solve OCO-S with linear switching cost have unbounded competitive ratios. Next, we show that an algorithm that 
  chooses $s(t) = \frac{n(t)}{\alpha^{1/4}}$ has a competitive ratio of $O(\alpha^{1/4})$ and a matching lower bound for any algorithm in the class $s(t) = \frac{n(t)}{\alpha^{\gamma}}$ for any $\gamma>0$. This 
  establishes that the considered problem in the linear swiching cost case has a very different characterization than OCO-S with linear switching cost where constant 
  competitive ratios are achievable in $1$-dimension. 
  
 The dependency of our results on $\alpha$ compared to OCO-S \eqref{eq:oco} with linear switching cost where they are independent is  because with 
 OCO-S one can equivalently write the cost in \eqref{eq:oco} as ${\hat f}_t(.)+c(x(t),x(t-1))$, where 
  ${\hat f}_t(.) = \frac{f_t(.)}{\alpha}$ that is still convex.
  Essentially, $\alpha$  can be absorbed within the convex function $f_t$ revealed at time $t$ itself in OCO-S \eqref{eq:oco} with linear switching cost.

\item When the switching cost is quadratic, we propose a simple online
  algorithm, that at any time slot $t$ chooses the number of servers
  to operate as $s(t) = \min\{\beta \sqrt{\frac{n(t)}{\alpha}}, n(t)\},$ where $\beta \ge 1$ (to be chosen), and uses multi-server shortest
  remaining processor time (SRPT) algorithm for scheduling.\footnote{When $\alpha<1$, replace $\alpha$ with $1$ in $s(t)$ definition.}  
  We show
  that this algorithm has a competitive ratio of at most $20$ for an appropriate choice of $\beta$, irrespective of the value of $\alpha$. 
  
  Thus, the linear switching cost case and quadratic switching cost case are fundamentally different in terms of $\alpha$ dependence.
  The main reason for this difference is that with the quadratic switching cost, $\opt$ cannot avoid paying a comparable 
  switching cost to the proposed algorithm, while in the linear switching cost case $\opt$ can use its extra information about the future job arrivals and adjust its 
  speed choices so as to pay an orderwise less switching cost than online algorithms when $\alpha >1$.

  Compared to prior work on OCO-S for the
  quadratic switching cost \cite{chen2018, goel2019online,
    goel2019beyond} our analysis is novel, and uses the primal-dual
  technique to obtain the competitive ratio guarantee. In prior work
  on OCO-S, mainly a potential function approach is used that works well when there is no memory in the cost function.
  
  
  The primal-dual technique is a
  very versatile recipe and has been applied for related problems, however, mostly for speed scaling \cite{nguyen2013lagrangian, devanur2017primal,
    vazebook} and not with switching cost. Applying primal-dual technique for the considered problem is more challenging because of the unbounded memory as well as a natural requirement that we need to enforce called the {\it work neutrality constraint}, and the switching cost. Work neutrality essentially means that the total work done by an algorithm until any time is at most the total work brought into the system by that time, precluding server idling. Since we are considering worst case input,  $\opt$ knowing the future input might idle some servers in hope of saving on the switching cost later, a behaviour which no online algorithm can match. More details as to why enforcing work neutrality constraint is necessary are presented in Remark \ref{rem:idle}.
    \newline
 At this point it is worth recalling that for OCO-S with quadratic switching cost,  the competitive ratio of any online algorithm is unbounded when cost functions $f_t$'s are just convex and not strongly convex. The input that achieves this is $f_t(x)=0, \forall \ x$ for $t=1,\dots, T-1$ and $f_T(x) = (x-1)^2$. Now if any online algorithm 
 $\cA$ chooses an action $x_t\ne 0$ for some $t$, then the input stops at time $t$, and $\cA$ has non-zero cost \eqref{eq:oco} while the cost of $\opt$ is $0$. Otherwise, if $\cA$ keeps $x_t=0$ for $t=1,\dots, T-1$, then it has to pay a constant cost at time $T$. $\opt$ on the other hand knowing the input, chooses to move towards $1$ slowly 
 from $t=0$ to $t=T$, paying a cost of $O(1/T)$. Thus, the competitive ratio of $\cA$ is at least $\Omega(T)$. Choosing $T$ large, the competitive ratio becomes unbounded. For the considered problem \eqref{eq:introocoflowtime1}, the work neutrality constraint essentially avoids this pitfall by not allowing $\opt$ to operate non-zero number of active servers when there is no work in the system.

\item When the switching cost is quadratic and all jobs are available
  at time $0$, we also characterize an optimal algorithm when $s(t)$ is allowed to be non-integral. 
\end{itemize}
\item Stochastic Input\\ In the stochastic setting, we assume that
  jobs arrive as per a Poisson process with rate~$\lambda$ and job
  sizes are independent and identically distributed (i.i.d.)
  exponential with unit mean. The cost objective is naturally modified
  to be the long run time average of the system occupancy, plus the
  long run rate at which switching cost is accrued.
  \begin{itemize}
  \item For both linear and quadratic switching costs, we find that a
    simple algorithm that operates as many active servers as
    outstanding jobs ($s(t) = n(t)$) is
    $(1+2\alpha)$-competitive. Interestingly, this algorithm (and
    consequently also the optimal algorithm) has a cost that is
    $O(\lambda).$
  \item We also consider a single server speed-scaling variant of the
    above model, wherein the speed of a single server can be modulated
    with a quadratic switching cost. Interestingly, we find that in
    this case, overall cost that grows $o(\lambda)$ is attainable with
    a policy that operates at a single speed, but only after letting
    the buffer build to a certain threshold.
  \end{itemize}
  
\end{enumerate}

%% file: ProblemFormulation.tex
\section{Worst-Case Input}
Consider a discrete time system where each slot width is normalized to unity.
Let the set of jobs that arrive over time be $\cJ$. For any job $j\in \cJ$, its arrival time is $a_j$ while its size is $w_j \in \bbZ^+$, which is revealed on arrival. 
We will assume identical job sizes for our analysis, i.e. $w_j = w_k, \ \forall j\ne k$, and let the job size be $w$. We let all job arrivals and departures to happen at the start and end of a slot, respectively. 
From here on, time means a time slot. Following the classical work \cite{lin2012online},  we assume that there is an unlimited number of servers available, each with fixed speed of $1$, and any job at any time instant can be processed by at most one server and each server can process at most one job at any time. 
Job $j$ departs as soon as $w_j$ amount of work has been completed for it possibly over different servers. Let the departure time of job $j$ be $d_j$. Then the flow time for job $j$ is $f_j = d_j-a_j$, and 
the total flow time $F$ is $\sum_{j\in \cJ} f_j$. Note that if $n(t)$ is the number of outstanding jobs at time $t$, then $F = \sum_{t} n(t)$.

Let the number of servers being used (called {\bf active}) at time $t$ be $s(t)$. Thus, at time $t$, $s(t)$ distinct jobs are getting processed in $s(t)$ servers at speed $1$. Increasing the number of servers decreases the flow time, however, changing the number of active servers across time is costly and the switching cost is modelled as $c(s(t), s(t-1))$ that is generally an increasing function of $|s(t)- s(t-1)|$. Moreover, there is an energy cost needed to run $s(t)$ servers at time $t$ that is modelled as $\theta \cdot s(t)$ for some $\theta>0$. 

Then the problem is to minimize the sum of the flow time, the switching cost, and the energy cost given by 
\begin{equation}\label{eq:prob1} \min_{s(t)}  \ \ \  \sum_{t} n(t) + \alpha \sum_{t}  c(s(t), s(t-1)) + \sum_{t} \theta \cdot s(t),
\end{equation}
where $\alpha$ is the tradeoff factor between the importance given to the flow time and the switching cost.
We consider two switching cost functions, linear $c(s(t), s(t-1)) = |s(t)-s(t-1)|$ and quadratic $c(s(t), s(t-1)) = (s(t)-s(t-1)^2$. For reasons described in Remark \ref{rem:idle}, in the quadratic case we need to enforce
the {\it work neutrality} constraint that the amount of work done by time $t$ cannot be more than the total amount of work that has arrived till then, i.e. $\sum_{t\le t'} s(t) \le \sum_{t\le t'} w(t)$ for all $t'$, where $w(t) = \sum_{j\in \cJ, a_j =t} w_j$ is the sum of sizes of jobs that arrive at time $t$. Essentially, servers are not allowed to idle in order to save on future switching costs. If  work neutrality is not enforced $\opt$ has an unfair advantage (as shown in Remark \ref{rem:idle}) over all online algorithms. For the ease of exposition, we assume that work neutrality is enforced throughout and point out in Remark \ref{rem:idlelinear} that it is not needed for the linear switching cost case.
With the work neutrality constraint, for any algorithm the sum of the number of active servers across all time is equal to the total work, i.e., $\sum_{t} s(t) = \sum_{j\in \cJ} w_j$. 

Hence we consider the following objective function.

\begin{equation}\label{eq:prob} \min_{s(t)}   \sum_{t} n(t) + \alpha \sum_{t}  c(s(t), s(t-1)),
\end{equation}
under the work neutrality constraint. 
\begin{remark} Typically in scheduling problems work neutrality constraint does not have to be explicitly written since violating it only entails larger flow time or energy cost. For problem \eqref{eq:prob}, however, it is critical to enforce it since violating it can lead to smaller switching cost compared to increased flow time and can be exploited by an optimal offline algorithm $\opt$. See Remark  \ref{rem:idle}
\end{remark}

\begin{remark} We are modelling the energy cost as linear in the number of servers used since each server is of fixed speed and at most one job can be processed on one server. In speed scaling models \cite{bansal2009speed, vaze2020multiple} with a single or multiple servers, each server is of variable speed and choosing speed $s$ the instantaneous power cost is modelled as $P(s)$, where $P(.)$ is a strictly increasing convex function.
\end{remark}
We consider the online setting where an online algorithm $\cA$ at time $t$ is only aware of the jobs that have arrived till time $t$, and makes its 
decisions about $s(t)$ to solve \eqref{eq:prob} using only that. Even $|\cJ|$ is not known to $\cA$. To quantify the performance of $\cA$, we use the metric of competitive ratio that is defined as follows.

We represent the optimal offline algorithm (that knows the entire job arrival sequence in advance) as $\opt$. 
Let $n(t)$ ($n_o(t)$) and $s(t)$ ($s_o(t)$) be the number of outstanding jobs with an online algorithm $\cA$ ($\opt$), and the number of servers used by an online algorithm $\cA$ ($\opt$) at time $t$, respectively. For Problem \eqref{eq:prob}, we will consider the metric of competitive ratio which for an algorithm $\cA$ is defined as 
\begin{equation}\label{defn:cr}\mu_\cA  = \max_\sigma \frac{\sum_{t} n(t) + \alpha \sum_{t} c(s(t), s(t-1))}{\sum_{t} n_o(t) + \alpha \sum_{t}  c(s_o(t), s_o(t-1))},
\end{equation}
where $\sigma$ is the input sequence consisting of jobs set $\cJ$. Since $\sigma$ is arbitrary, we are not making any assumption on the job arrival times, or $|\cJ|$ the total number of jobs . The goal is to derive online algorithms with small competitive ratios.

Two specific switching cost functions that have been considered extensively in literature are i) linear $c(s(t),s(t-1)) = | s(t)-s(t-1)|$ and ii) quadratic $c(s(t),s(t-1)) = (s(t)-s(t-1))^2.$ We consider both the cases and begin with the linear case as follows.
 
\subsection{Linear Switching Cost}\label{sec:linear}
With linear switching cost, problem \eqref{eq:prob}  simplifies to
\begin{equation}\label{costlinear}
\min_{s(t)} C_{\cA}, \quad  C_{\cA}=  \sum_{t} n(t) + \alpha \sum_{t}  |s(t)- s(t-1)|.
\end{equation} 
We first encounter the easy case of $\alpha \le 1$ and then present the challenging one when $\alpha > 1$.
\subsection{$\alpha \le 1$} 
Consider an algorithm $\cA_f$ that chooses $s(t) = n(t)$ for each time slot $t$. 

\begin{lemma}\label{lem:alphale1}
  $\cA_f$ is $2$-competitive, and the competitive ratio is tight.
\end{lemma}
Essentially, when $\alpha \le 1$, the cost of $\cA_f$ is $\le 2  \sum_t n_{\cA_f}(t)$, while $ \sum_t n_{\cA_f}(t)$ is a lower bound on the cost of $\opt$. 
All the missing proofs are provided in the Appendix.

\subsection{$\alpha > 1$} The $\alpha>1$ case is more interesting and challenging case compared to $\alpha \le 1$. 
With $\alpha>1$, the most natural algorithm to minimize \eqref{costlinear} inspired from prior work on OCO-S  \cite{lin2012online,
  lin2012dynamic, andrew2013tale, bansal20152,
  DBLP:journals/corr/abs-1807-05112,chen2018, argue2020dimension}
 is to choose $s(t)$ so as to balance the two instantaneous costs, flow-time $n(t)$ and switching cost $\alpha |s(t)-s(t-1)|$ which takes either of the two forms $s(t) = \frac{n(t)}{\alpha}$ or $|s(t)-s(t-1)| = \frac{n(t)}{\alpha}$. \footnote{Take ceiling if not an integer.}
We next show that for the considered problem an algorithm $\cA_{bal-1}$ that chooses 
$s(t) = \frac{n(t)}{\alpha}$ and $\cA_{bal-2}$ that chooses 
$|s(t)-s(t-1)| = \frac{n(t)}{\alpha}$ has an unbounded competitive ratio. 

\begin{lemma}\label{lem:balanced}
$\cA_{bal-1}$ and $\cA_{bal-2}$ have unbounded competitive ratios.
\end{lemma}


Thus, we need {\it non-balanced} algorithms. Towards that end, we first consider the simpler setting of all jobs arriving at time $0$ with equal size $w$. 
With $\sfN$ total jobs that arrive at time $0$, we next show an algorithm that uses $s(t) = \Theta\left(\frac{\sfN}{\alpha^{1/2}}\right)$ from time $0$ until all jobs are finished subject to satisfying the work neutrality constraint is optimal. 
\begin{lemma}\label{lem:linearopt}
When all $\sfN$ jobs arrive at time $0$, algorithm $\cA_{sqrt}$ that chooses $s(t) = \frac{\sqrt{\sfN(\sfN-1)}}{2\alpha^{1/2}}$ for as long as possible to satisfy the work neutrality constraint and thereafter uses $s(t)=n(t)$, is optimal, 
\end{lemma}

It turns out (Lemma \ref{lbcrlinearcase}) that the natural online extension of algorithm $\cA_{sqrt}$, called  $\cA_{sqrt-on}$, when jobs arrive over time, i.e. to use $s(t) =\max\left\{1, \frac{n(t)}{\sqrt{\alpha}}\right\}$, where $n(t)$ is the number of outstanding jobs at time $t$, has a competitive ratio of $\alpha^{1/2}$, however, smaller competitive ratios are achievable as we show next.

To improve the competitive ratio of  $\cA_{sqrt-on}$, we consider a more aggressive algorithm as follows.

{\bf Algorithm $\cA_{LG}$:} At time $t$, if the current active number of servers $s(t) > \frac{n(t)} {\alpha^{1/4}}$ do nothing. Otherwise, make $s(t) = \frac{n(t)} {\alpha^{1/4}}$, where $n(t)$ is the number of outstanding jobs at time $t$.
\begin{lemma}\label{lem:algLG}  $\cA_{LG}$ is at most $4\alpha^{1/4}$-competitive.
\end{lemma}
We provide the proof of Lemma  \ref{lem:algLG} here in the main body since it is elegant and novel compared to OCO-S with linear switching cost \cite{lin2012online,
  lin2012dynamic, andrew2013tale, bansal20152,
  DBLP:journals/corr/abs-1807-05112,chen2018, argue2020dimension} and uses a combination of two potential functions, one counting the number of active servers and other counting the outstanding amount of work between the algorithm and the $\opt$.
\begin{proof} Without loss of generality, we let $w=1$.
To bound the competitive ratio, we will first consider the case when the switching cost is counted only when number of active servers is increased, i.e., the cost of $\cA_{LG}$ is 
\begin{equation}\label{costlinearag}
C_{\cA_{LG}}^{\uparrow} =  \sum_{t} n(t) + \alpha \sum_{t}  (s(t)- s(t-1))^+,
\end{equation} 
where $(x)^+ = \max\{0,x\}$.
Given that $\cA_{LG}$ satisfies
the work neutrality constraint, and the switching cost is linear, the true cost $C_{\cA_{LG}}$ \eqref{costlinear}  is at most two times $C_{\cA_{LG}}^{\uparrow}$.

To bound the cost \eqref{costlinearag} of $\cA_{LG}$, 
consider the following potential function
\begin{equation}\label{}
\Phi(t) = \Phi_1(t) + \Phi_2(t),
\end{equation}
where 
\begin{equation}\label{}
 \Phi_1(t) = \alpha(-s(t)+ s_o(t)),
\end{equation}
where $s(t)$ and $s_o(t)$ are the number of active servers with $\cA_{LG}$ and $\opt$ at time $t$, respectively, and 
\begin{equation}\label{}
\Phi_2(t) = \alpha^{1/4} (W(t) - W_o(t)), 
\end{equation}
where $W(t) = w n(t)$ and $W_o(t)=w n_o(t)$ is the total work remaining with  $\cA_{LG}$ and $\opt$ at time $t$, respectively.

 Clearly, $\Phi(0) = 0$ and $\Phi(\infty) =0$. 
If at  time $t$ the number of active servers is increased by $\cA_{LG}$ 
 \begin{align}\nn
\Delta_1(t) &= \Phi_1(t) - \Phi_1(t-1), \\ \label{drift11}
& \le \alpha(-s(t) + s(t-1)) + \alpha | s_o(t) - s_o(t-1)|,
\end{align}
where $s(t) > s(t-1)$, and the first term is negative.
Otherwise, if at  time $t$ the number of active servers is decreased by  $\cA_{LG}$
 \begin{align}\nn
\Delta_1(t) &= \Phi_1(t) - \Phi_1(t-1), \\\label{drift12}
& \le \alpha(-s(t) + s(t-1)) + \alpha | s_o(t) - s_o(t-1)|,
\end{align}
where $s(t-1) < s(t)$, and hence the first term is positive.
When $s(t)=s(t-1)$, 
\begin{align}\label{drift13}
\Delta_1(t) &\le  \alpha | s_o(t) - s_o(t-1)|.
\end{align}

Moreover, for any $t$, 
 \begin{align}\nn
\Delta_2(t) &= \Phi_2(t) - \Phi_2(t-1), \\ \nn
&=\alpha^{1/4}(-s(t)) + \alpha^{1/4}(W_o(t)-W_o(t-1)), \\\label{drift2}
&\stackrel{(a)}\le -n(t) + \alpha^{1/4}n_o(t),
\end{align}
where $(a)$ follows by the choice of the number of active servers by $\cA_{LG}$ to be $s(t) \ge \frac{n(t)}{\alpha^{1/4}}$, while 
$(W_o(t)-W_o(t-1)) \le n_o(t)$ follows from the fact that each job can be served by at most one server at any time and the speed of each server is unity. Note that to get \eqref{drift2}, we are not enforcing work neutrality for the $\opt$.

Combining \eqref{drift11}, \eqref{drift12}, \eqref{drift13}, and \eqref{drift2},  we get for $\cA_{LG}$ at any $t$ \newline
$ \underbrace{n(t) + \alpha (s(t) - s(t-1))^+}_{C_{\cA_{LG}}^{\uparrow}} +\Delta_1(t)+\Delta_2(t) $
 \begin{align}\nn
&\stackrel{(a)}\le \alpha^{1/4}n_o(t) + \alpha| s_o(t) - s_o(t-1)| \\ \nn
 & \quad \quad + \b1_{\downarrow}(t)\alpha(-s(t) + s(t-1)), \\ \label{eq:finalruncond}
 & \stackrel{(b)}\le \alpha^{1/4} C_\opt(t) + \b1_{\downarrow}(t)\alpha(-s(t) + s(t-1)),
 \end{align}
 where in $(a)$ $\b1_{\downarrow}(t)$ is the indicator function that takes value $1$ for  slots $t$ where $\cA_{LG}$ decreases its active number of servers and $(-s(t) + s(t-1)) > 0$ and is $0$ otherwise, while in $(b)$ $C_\opt(t)$ is the cost \eqref{costlinear}  of $\opt$ at time $t$ since $\alpha >1$. 
 
 Recall that $C_{\cA_{LG}}^{\uparrow} = \sum_t n(t) + \alpha (s(t) - s(t-1))^+$ and $\Phi(0) = 0$ and $\Phi(\infty) =0$. Thus,
 summing \eqref{eq:finalruncond} over all $t$,  
 we bound the cost \eqref{costlinearag} of $\cA_{LG}$, $C_{\cA_{LG}}^{\uparrow}$, as 
  \begin{equation}\label{dummyyxyx1}
C_{\cA_{LG}}^{\uparrow} \le  \alpha^{1/4} C_\opt + \frac{1}{2}C_{\cA_{LG}}^{\uparrow},
\end{equation}
since $\sum_t \b1_{\downarrow}(t)\alpha(-s(t) + s(t-1))$ is at most half of the total switching cost out of the total cost \eqref{costlinearag} for $\cA_{LG}$ because of the linear switching cost and work neutrality constraint. Thus, \eqref{dummyyxyx1} gives
 \begin{equation}\label{eq:dummyxx1}
C_{\cA_{LG}}^{\uparrow} \le 2 \alpha^{1/4} C_\opt.
\end{equation}
From our earlier argument, the cost \eqref{costlinear} for $\cA_{LG}$
$$C_{\cA_{LG}}\le 2 C_{\cA_{LG}}^{\uparrow}$$ which together with \eqref{eq:dummyxx1} completes the  proof.
\end{proof}
\begin{remark}\label{rem:idlelinear} It is important to note that for both the $\alpha \le 1$ as well as the $\alpha>1$ case we have not used the work neutrality constraint in deriving the guarantees. Thus, even if $\opt$ keeps some servers idling anticipating more work to arrive in future in order to reduce its switching cost, the proposed algorithms can `catch up'. One might be tempted to argue that the competitive ratio being $4\alpha^{1/4}$ when $\alpha>1$ could be on account of non enforcement of  the work neutrality constraint on the $\opt$. The following lower bounds answer that in the negative.
\end{remark}

Next, we show that the competitive ratio of any online algorithm belonging to the class $s(t) = \frac{n(t)}{\alpha^\gamma}$ for any $\gamma \ge0$ is at least $\Omega(\alpha^{1/4})$. Note that choosing the number of active servers  $s(t)$ at any time naturally constrains the switching cost for any algorithm as well.
\begin{lemma}\label{lbcrlinearcase}
  Consider an online algorithm $\cA_\gamma$ that chooses $s(t) = \frac{n(t)}{\alpha^\gamma}$ for $\gamma\ge 0$. The competitive ratio of $\cA_\gamma$ is at least $\Omega(\alpha^{1/4})$.
\end{lemma}

{Discussion: } When $\alpha\le 1$, problem \eqref{costlinear} is simpler than the OCO-S with linear switching cost, and a trivial algorithm is $2$-competitive. In contrast when $\alpha > 1$, problem \eqref{costlinear} is fundamentally different than OCO-S with linear switching cost. In particular, as pointed out before, a simple balancing algorithm is $3$-competitive for OCO-S with linear switching cost \cite{bansal20152}, while as shown above the competitive ratio of any algorithm for problem \eqref{costlinear} depends on $\alpha$, and optimally grows as $\alpha^{1/4}$. Thus, even though there is 
great similarity between OCO-S and problem \eqref{costlinear}, the characterization is entirely different. This is essentially an artefact of the unbounded memory in the objective function with the considered problem.

\subsection{Quadratic Switching Cost}\label{sec:quad}
In this subsection, we consider Problem \eqref{eq:prob} with quadratic switching cost, i.e., 
 \begin{equation}\label{eq:probquad} \min_{s(t)}   \sum_{t} n(t) + \alpha \sum_{t}  (s(t)- s(t-1))^2.
\end{equation}
where  $n(t) = \max\left\{\sum_{\tau\leq t} \text{arr}_\tau - \sum_{\tau \leq t-1} \text{dep}_\tau,0\right\},$ where $\text{arr}_\tau$ and $\text{dep}_\tau$ are the number of jobs that arrive at the start of slot $\tau$ and depart at the end of slot $\tau$, respectively.


\begin{remark}\label{rem:idle} We now detail why we need the work neutrality assumption.  
If servers are allowed to idle, i.e., violate the work neutrality assumption, an offline algorithm gets an unfair edge over any online algorithm when the switching cost is {\bf quadratic}. Let $\alpha=1$, and consider for example the following input, where at time $t=T$, $T$ jobs of unit size arrive, and no jobs arrive before and after time $T$. 
An offline algorithm $O$ knowing this, will choose $s(t) = t$ for $t=1, \dots, T-1$, thus paying only a unit switching cost per unit time even when the switching cost is quadratic. At time $T$, when all the jobs arrive,  $O$ will choose $s(t) = T$, and finish all jobs in just one slot, and ramp down the speed as $s(T+t) = T-t$.
Thus, $O$ incurs a total switching cost of $2T$ and the flow time cost of $T$, resulting in a total cost of $3T$. In comparison, any online algorithm $\cA$ cannot use $s(t) >0$ until $t=T-1$ since otherwise the input can be such that no jobs arrive ever, making the competitive ratio of $\cA$ arbitrary large. Thus, $\cA$ can only choose $s(t)>0$ for $t \ge T$, in which case its total cost is $\Omega(T^{3/2})$. Thus, if we do not make this critical modelling assumption, no algorithm can be 
competitive while choosing $T$ large. 

Practically also, if servers are allowed to idle, under the uncertain future, any algorithm might incur a large running cost anticipating the arrival of new jobs, compared to an offline algorithm which knows the exact input. 
\end{remark}
\vspace{-0.1in}
 \subsection{Algorithm}
 For solving \eqref{eq:probquad}, we consider the following algorithm, called $\textsf{alg}$ from here on. 
 {\bf The choice of number of active servers} $s(t) = \min\{\beta \sqrt{\frac{n(t)}{\alpha}}, n(t)\},$ where $\beta \ge 1$ to be chosen later. If $s(t)$ is not an integer, take the ceiling. When $\alpha<1$, replace $\alpha$ with $1$ in $s(t)$ definition.
 
 {\bf Scheduling:} Multi-server SRPT, i.e., for the chosen number of active servers $s(t)$, process the $s(t)$ jobs with the shortest remaining processing time/job size. 
 If two jobs are of same size, process them in order of their arrival times. Break ties arbitrarily if the arrival time is also the same.
 Note that the execution of multi-server SRPT requires both preemption and job-migration, i.e. a job will be processed on possibly different servers at different instants of time.
 
 \begin{remark}  By definition, $\textsf{alg}$ satisfies the work neutrality constraint.
 \end{remark}
 
  \begin{remark}  As long as $\beta \sqrt{\frac{n(t)}{\alpha}} = \min\{\beta \sqrt{\frac{n(t)}{\alpha}}, n(t)\}$, $\textsf{alg}$ chooses speed at time $t$ such that the 
switching cost is at most $\beta^2$ times the number of remaining jobs at time $t$. Thus, $\textsf{alg}$ tries to relate the two costs of \eqref{eq:probquad}. This 
is similar to algorithms proposed for the OCO-S problem with quadratic switching cost
\cite{chen2018, goel2019beyond}, however, the analysis of $\textsf{alg}$ is entirely different since the 
optimal choice at time $t$ for minimizing the flow time in \eqref{eq:probquad} depends on the past choices of the number of active servers.
 \end{remark}

For $\textsf{alg}$, it is easy to upper bound the total cost \eqref{eq:probquad} as follows. 
\begin{lemma}\label{eq:costalg} The cost \eqref{eq:probquad} of $\textsf{alg}$ is at most 
$(1+2\beta^2) \sum_t n(t),$
where $n(t)$ is the number of remaining jobs at time $t$ with the $\textsf{alg}$.
\end{lemma}

It might appear that there is significant overcounting in the proof of Lemma \ref{eq:costalg} to upper bound the switching cost, however, one can construct 
input instances where the counting is in fact tight.

Following is the {\bf main result} of this section.
 \begin{theorem}\label{thm:main} The competitive ratio of $\textsf{alg}$ for solving \eqref{eq:probquad} is at most $20$.
 \end{theorem}
 The proof is rather long and is provided in the Appendix.
 Here we only give a proof sketch as follows.
 Problem \eqref{eq:probquad} can be equivalently written as 
 \begin{equation}\label{eq:probeqmainbody}
\begin{aligned}
& \underset{s(t)}{\min}
& & \sum_j \left(\sum_{t=a_j}^{d_j} s_j(t) \right) \frac{d_j-a_j}{w_j}+ \alpha \sum_{t}  (s(t)- s(t-1))^2 \\
& \text{}
& & \begin{aligned} \sum_{t=a_j}^{d_j} s_j(t)  & \ge w_j, \ \forall \ j, \quad  \sum_{\tau \le t} s(\tau) \le \sum_{\tau\le t} w(\tau) \ \forall \ t,\\
 s_j(t) &\geq 0 \ \forall \ t, j, \quad   \sum_{j} s_j(t) = s(t) \ \forall \ t,\\
 \end{aligned}
\end{aligned}
\end{equation}
where $s_j(t)=1$ if  job $j$ is being processed at time $t$ and $0$ otherwise, and $s(t)$ is the total number of active servers at time $t$, and the constraint $\sum_{t=a_j}^{d_j} s_j(t) \ge w_j$ implies that job $j$ is eventually completed.
The constraint $\sum_{\tau \le t} s(\tau) \le \sum_{\tau\le t} w(\tau) \ \forall \ t,$ corresponds to the work neutrality constraint and for brevity we denote it as 
$\mathsf{acc}(t)\le D(t)$ hereafter, where $\mathsf{acc}(t)=\sum_{\tau \le t} s(\tau)$, while $D(t)= \sum_{\tau\le t} w(\tau)$.

The Lagrangian dual function associated with \eqref{eq:probeqmainbody} is \newline 
$
 \min_{s(t), \ d_j, \ \mathsf{acc}(t)\le D(t) } \cL(\lambda_j, \mu_j, \nu_t)=\min_{s(t), \ d_j, \ \mathsf{acc}(t)\le D(t)}$
\begin{align}\nn&   \quad  \sum_j \left(\sum_{t=a_j}^{d_j} s_j(t)  \right) \frac{d_j-a_j}{w_j} + \alpha \sum_{t}  (s(t)- s(t-1))^2 \\ \label{eq:dummymain} 
& \quad + \sum_j \lambda_j \left(w_j - \sum_{t=a_j}^{d_j} s_j(t)\right)   + \sum_j \mu_j s_j(t) + \sum_t \nu_t (\sum_{j} s_j(t) - s(t)),
\end{align}
where  $\lambda_j, \mu_j,$ and $\nu_t$ are the dual variables corresponding to $\sum_{t=a_j}^{d_j} s_j(t)   \ge w_j$, $s_j(t) \geq 0 \ \forall \ t, j$, and $\sum_{j} s_j(t) = s(t)$, respectively, and {\bf we have kept the work neutrality constraint as is in \eqref{eq:dummymain}.} Keeping the work neutrality constraint as it is is in fact critical for the analysis. Instead if it is included inside the Lagrangian, that results in a trivial lower bound that is not useful. This feature makes the analysis sufficiently novel compared to usual application of primal-dual scheme for similar problems.

 {\bf Choosing the dual variable $\lambda_j$} Given the $\textsf{alg}$, choose $\lambda_j$ such that
$\lambda_j w_j = \Delta_j F_{\textsf{alg}}$, where $\Delta_j F_{\textsf{alg}}$ is the increase in the total flow time of the $\textsf{alg}$ because of arrival of job $j$ with size $w_j$ at time $a_j$, disregarding jobs arriving after time $a_j$.

For this choice of $\lambda_j$, we obtain the following important structural result for $\textsf{alg}$. 
\begin{lemma}\label{lem:lambdajtimemainbody}
\begin{equation}\label{eq:lambdajtconnectionmainbody}
\lambda_j - \left(\frac{t-a_j}{w_j}\right) \le \frac{3}{\beta} (\alpha n(t))^{1/2}
\end{equation} for all $t$ and $\beta \ge 1$, where 
$n(t)$ is the number of remaining jobs with  $\textsf{alg}$ at time $t$. 
\end{lemma}

Combining Lemma \ref{lem:lambdajtimemainbody} and invoking work neutrality, we get the following lower bound on the Lagrangian  in \eqref{eq:dummymain}
\begin{align}\label{eq:lbdual3mainbody} \ge \sum_t n(t)\left(1-\frac{9}{4\beta^2}\right),
\end{align}
where $n(t) $ is the number of remaining jobs  with $\textsf{alg}$ at time $t$.

From the weak duality, the primal value of \eqref{eq:probquad} is at least as much as the value of any dual feasible solution. Thus, the competitive ratio of 
any online algorithm $\cA$ is at most the ratio of an upper bound on the cost of $\cA$ and a lower bound on the dual feasible solution. Hence combining  \eqref{eq:lbdual3mainbody} and Lemma \ref{eq:costalg}, we get that the competitive ratio of $\textsf{alg}$ is at most 
$$\mu_{\textsf{alg}} \le \frac{1+2\beta^2}{\left(\frac{4\beta^2-9}{\beta}\right)}.$$
For $\beta^2=3$, we get $\mu_{\textsf{alg}} \le 7\times 4= 28,$
while optimizing over $\beta$, we get that the optimal $\beta = 2.177$ and the competitive ratio is $19.95$.
This completes the proof of Theorem \ref{thm:main}.

  
{\it Discussion:} Theorem \ref{thm:main}  shows that $\textsf{alg}$ that tries to balance the flow time and switching cost of \eqref{eq:probquad} has a constant competitive ratio. This balancing idea is also integral to algorithms analyzed for OCO-S with quadratic switching cost problem, however, the analysis of $\textsf{alg}$ is entirely different 
 compared to OCO-S with quadratic switching cost problem. The usual technique to analyze OCO-S with quadratic switching cost problem is to consider a potential function $\Phi_1(x(t), x^o(t))  = || x(t) - x^o(t)||^2$, where $x(t)$ and $x^o(t)$ are the actions chosen by an algorithm and the $\opt$ at time $t$, respectively,  and critically use the property that the optimal choice to minimize the cost $f_t(.)$ in
 \eqref{eq:oco} is $x_t^\star$ that depends only on $f_t$. Since this is not the case with \eqref{eq:probquad}, we deviate from the potential function approach and consider the primal-dual scheme for analysis. 
 
 The main advantage of the primal-dual scheme is that there is no need to characterize the structure or any  other property of the $\opt$ directly, which is anyway not available for  \eqref{eq:probquad}.  
 By exploiting the monotonicity property of $\textsf{alg}$ (an algorithm is defined to be {\bf monotone} if $n^{\cI}(t) \le n^{\cI'}(t)$, where $n^{\cI}(t)$ is the number of remaining jobs at time $t$ with job arrival instance $\cI$ and $\cI\subseteq \cI'$)
 and the fact that any new job does not preempt an existing job (which is true since all job sizes are identical) for $\textsf{alg}$, we derive an upper bound on the primal cost and a lower bound on the dual cost with $\textsf{alg}$, which together with the weak duality implies the competitive ratio bound. 
 
 The analysis we use is similar to \cite{nguyen2013lagrangian} that considered a flow-time +energy minimization problem but without any switching costs or discrete number of active servers constraint or the work neutrality constraint. Compared to usual primal-dual technique analysis, e.g. \cite{nguyen2013lagrangian}, the most challenging new part is the enforcement of the work neutrality constraint as described in Remark \ref{rem:idle}. 
 As far as we know primal-dual analysis has not been applied for solving the OCO-S with quadratic switching cost problem \eqref{eq:oco}  or its variants, and we believe it is going to be quite useful for other OCO-S problems as well. It is important to note that our analysis does not work for arbitrary objective
 convex functions $f_t's$ as described in OCO-S with quadratic switching cost, thus our work critically exploits the specific objective function of \eqref{eq:probquad}.

 

Next,   we solve the simpler case of Problem \ref{eq:probquad} when all jobs arrive at the same time, optimally, as long we do not enforce the integrality constraint on the number of active servers $s(t)$. 

To solve \eqref{eq:probquad} exactly when all jobs arrive at the same time, consider the following optimization problem. 
Let there be~$n$ unit sized jobs available at time~$0$, and  problem $\bP_H$ is to minimize the flow time plus the quadratic switching cost over a
horizon of~$H$ time slots, i.e., all $n$ jobs should be completed by $H$ time slots.
$\min_H \bP_H$, when all job sizes are unity. Thus if we can solve $\bP_H$ for any $H$ (structurally) we can solve \eqref{eq:probquad} as well.
To solve $\bP_H$, let~$s(t)$ denote the number of servers
active in time slot~$t$ equivalently, $s(t)$ denotes the number of
jobs completed in time slot~$t.$

Note that with $\bP_H$, the total flow time equals~$(n-s(1)) + (n-s(1)-s(2))+\cdots$. Thus, $\bP_H$ is to solve
\begin{align}
  \min \quad Hn - &\sum_{i=1}^{H+1}s(i)(H+1-i) + \sum_{i = 1}^{H+1}(s(i) - s(i-1))^2 \nonumber \\
  \text{ s.t. } & s(0) = s(H+1) = 0,\\  \label{opt:burst}
  &s(i)\geq 0 \ \forall\ i, \ \sum_{i=1}^H s(i) = n. 
  \end{align}
Letting  optimization variables $\{s(i)\}$ be non-negative reals, the above optimization is convex, and admits an
explicit solution characterization given as follows.
\begin{lemma}\label{lem:conx}
  The optimal solution corresponding to the
  optimization~\eqref{opt:burst} is characterized as
  follows. 
  Let $$\lambda^\star = \frac{24n}{H(H+1)(H+2)} +
  \frac{(3H-2)(H+1)}{(H+2)}.$$ The optimal solution is:
  \begin{align}\nn
    s(1) &= \frac{1}{4}\left(n\lambda^\star - \frac{H(H-1)}{2} \right),\\ 
   \label{opt:soln} s(i)
    - s(i-1) &= s(1) -
    \frac{\lambda^\star(i-1)}{2}-\frac{(i-1)(H+1-\nicefrac{i}{2})}{2} \quad
    (i \geq 2).
  \end{align}
\end{lemma}
The proof follows from a straightforward application of the KKT
conditions and hence omitted.  Interestingly, the optimal speed profile is symmetric
around the mid-point of the horizon, and appears `concave' (i.e., it
has diminishing increments until the mid-point of the horizon, and
increasing decrements after that point). Since the solution provided
in Lemma~\ref{lem:conx} is in closed form, problem \eqref{eq:probquad}
can be solved by optimizing over a single variable $H$.


 

 

%% file: stochastic.tex
\section{Stochastic setting}
\label{sec:stochastic}

In this section, we consider a stochastic variant of the flow time
plus switching cost minimization problem analysed in the preceding
sections. Our goal as before is to design algorithms that are constant
competitive (relative to the optimal online algorithm). Interestingly,
we find that the nature of constant competitive algorithms in the
stochastic setting (with Poisson arrivals) differs from those we
established as competitive in the worst case setting (where burst
arrivals are possible). Finally, we also consider a single server
speed scaling variant of our model in the stochastic setting, where
the aggregate service rate is not constrained by the number of
outstanding jobs, and show that this relaxation results in a
significant cost reduction in heavy traffic.

\subsection{Model}

We begin by describing the system model for the stochastic setting. We
consider a continuous time model over an infinite horizon, and
consequently, redefine the cost metric \eqref{eq:probquad} in terms of
long run time averages of the system occupancy and switching cost.

Jobs arrive as per a Poisson process having rate~$\lambda.$ Job sizes
are i.i.d. and exponentially distributed with unit mean. Let $n(t)$
denote the number of jobs in the system at time~$t,$ and $s(t)$ denote
the aggregate service rate at time~$t,$ which is assumed to be
piecewise constant across time. Let $\mathrm{SC}(t)$ denote the total
quadratic switching cost associated with the process $\{s(\cdot)\}$
over the interval~$[0,t].$\footnote{Formally, $\mathrm{SC}(t)$ is the
  squared-total-variation associated with $s(\cdot)$ over the
  interval~$[0,t],$ i.e., $$\mathrm{SC}(t):= \sup_{P \in
    \mathcal{P}([0,t])} \sum_{i=0}^{n_P-1} |s(x_{i+1})-s(x_i)|^2,$$
  where $\mathcal{P}([0,t])$ is the set of
  partitions~$P=\{x_0,x_1,\ldots,x_{n_P}\}$ of $[0,t].$} The cost
metric we consider is the almost sure limit~$C$ of
\begin{equation}
  \label{eq:stochastic_obj_gen}
  \lim_{t \ra \infty} \left[\frac{1}{t}\int_0^t n(s) ds + \alpha
    \frac{\mathrm{SC}(t)}{t} \right].
\end{equation}
Specifically, we restrict attention to policies which admit a constant
almost sure limit corresponding to~\eqref{eq:stochastic_obj_gen}. Note
that the first term above can be interpreted as a long run time
average associated with the flow time metric (first term in
\eqref{eq:probquad}). The second term is an analogous long run time
average associated with the switching cost (second term in
\eqref{eq:probquad}); it may also be interpreted as the \emph{switch
  cost rate}, i.e., the rate at which switching cost is accrued in the
long run. Of course, $\alpha > 0$ denotes the weight parameter as
before. Finally, we note that our presentation in this section is
restricted to quadratic switching costs due to space constraints; our
results under quadratic switching costs generalize easily to linear
switching costs case as well (unlike the worst case setting, where the
analysis for these two costs structures is very different).

We consider two different models for the permissible choices of
aggregate service rate~$s(t).$\\
%
\noindent {\bf Multiserver model:} Here $s(t)$ is constrained to lie
in the set $\{0,1,2,\cdots,n(t)\}.$ This model is consistent with the
model we have employed in the worst case setting; with each server
assumed to have a unit speed, the aggregate service rate $s(t)$ equals
the number of active servers.\\
\noindent {\bf Single server speed-scaling model:} In this
generalization, $s(t)$ can take any value in $[0,\infty)$ when $n(t) >
  0,$ but $s(t) = 0$ when $n(t) = 0$ (to ensure \emph{work
    neutrality}).
  
  Finally, it is instructive to represent the cost metric~$C$ under
  Markovian policies. Given the memorylessness of the inter-arrival
  times and service requirements, under a Markovian policy, the
  aggregate service rate is purely a function of the number of jobs in
  the system at any time. Specifically, let~$\mu_i$ denote the service
  rate when there are~$i$ jobs in the system. Note that any Markovian
  policy induces a (time reversible) birth-death Markov chain on the
  number of jobs in the system; the time averages in
  \eqref{eq:stochastic_obj_gen} being meaningful when this chain is
  positive recurrent. In this case, let $N$ denote the steady state
  number of jobs in the system, and $\pi$ denote its distribution (i.e., the
  stationary distribution). Then
  \begin{align*}
    C &= \sum_{i=0}^{\infty} i \pi_i + \alpha \sum_{i=0}^{\infty} \lambda \pi_i
  (\mu_i-\mu_{i+1})^2 + \alpha \sum_{i=i}^{\infty} \mu_i \pi_i
    (\mu_i-\mu_{i-1})^2, \\
    &= \Exp{N} + 2 \alpha \sum_{i=0}^{\infty} \lambda \pi_i
  (\mu_i-\mu_{i+1})^2.
  \end{align*}

\subsection{Multiserver model}

We begin by considering the multiserver model described above, which
is simply the stochastic variant of the worst case model considered in
Section~\ref{sec:quad}. Consider algorithm $\alg_1,$ which sets $\mu_i
= i$ for all~$i$ (this means $s(t) = n(t)$ in the language of the
preceding sections).
\begin{lemma}
  \label{lemma:multiserver}
  The cost under $\alg_1$ is given by $C_{\alg_1} =
  \lambda(1+2\alpha).$ Moreover, $\alg_1$ is
  $(1+2\alpha)$-competitive.
\end{lemma}
\begin{proof}
  It is easy to see that under $\alg_1,$ $\pi_i = e^{-\lambda}
  \frac{\lambda^i}{i!}.$ It then follows that $\Exp{N} = \lambda$ and
  the switching cost rate under $\alg_1$ equals~$2 \alpha \lambda.$

  Since $s(t)$ can be at most $n(t),$ for any algorithm, it is easy to
  see that the long run average number in system is at-least that
  under $\alg_1.$ In other words, the cost associated with the optimal
  policy is at-least~$\lambda,$ which implies that $\alg_1$ is
  $(1+2\alpha)$-competitive.
\end{proof}

To summarize, we see that the continuous time stochastic variant of
the flow time plus (quadratic) switching cost problem admits a simple
constant (only depends on $\alpha$) competitive algorithm. In the worst-case input, we 
were able to show a $20$-competitive algorithm independent of $\alpha$. This 
difference is primarily on account of the difference in the two models, and where one 
does not subsume the other.

\begin{remark}
It follows from Lemma~\ref{lemma:multiserver} that in the multiserver
model, $\lambda$ is a lower bound on the cost of the optimal
algorithm. In other words, a cost that is $o(\lambda)$ is
unattainable. However, we will see that this is not the case in the
(more general, from a scheduling standpoint) single server
model. Intuitively, this is because it is possible to operate the
system such that there is a lot less switching (incurring a
$o(\lambda)$ switch cost rate) in the single server model, while still
operating at a high enough service rate to keep the holding cost small
(also $o(\lambda)$).
\end{remark}

\subsection{Single server speed-scaling model}

Next, we consider the single server speed-scaling model; our goal
being to contrast the scaling of cost with the multiserver model. We
find that in heavy traffic (i.e., when $\lambda$ is large), the
additional flexibility in the single server model enables a
significant reduction in cost.

We first consider the following adaptation of $\alg_1$ for the single
server model, and show that it is optimal in light traffic (as
$\lambda \da 0)$. Specifically, under~$\alg_2,$ $\displaystyle \mu_i =
\frac{1}{\sqrt[3]{4\alpha}}\ i.$
\begin{lemma}
  \label{lemma:singleserver_propspeed}
  The cost under $\alg_2$ is given by $C_{\alg_2} = \frac{3}{2}
  \sqrt[3]{4\alpha}\ \lambda.$
\end{lemma}
\begin{proof}
  Proof follows along similar lines as that of
  Lemma~\ref{lemma:multiserver}.
\end{proof}
\begin{lemma}
  \label{lemma:singleserver_lighttraffic_lb}
  Let $C_{\opt}$ denote the cost under the optimal Markovian
  policy. Then~$$\liminf_{\lambda \da 0} \frac{C_{\opt}}{\lambda} \geq
  \frac{3}{2} \sqrt[3]{4\gamma}.$$ It follows that~$\alg_2$ is
  light-traffic optimal.
\end{lemma}
\begin{proof}
  Under any Markovian policy~$\nu = (\mu_1,\mu_2,\ldots),$ we note
  that the cost admits the following lower bound. $$C_\nu \geq \lambda
  \left(\frac{1}{\lambda + \mu_1} + \frac{2\alpha \mu_1^2}{\lambda +
    \mu_1} \right).$$ This lower bound follows from an elementary
  sample path argument, wherein arrivals that take place when there is
  already a job in service are simply dropped, resulting in a 2-state
  (ON-OFF) Markovian description of the system occupancy. It now
  follows that $$C_{\opt} \geq \lambda \left(\frac{1}{\lambda +
    \beta^*(\lambda)} + \frac{2\alpha (\beta^*(\lambda))^2}{\lambda +
    \beta^*(\lambda)} \right),$$ where $\beta^*(\lambda)$ is the
  unique minimizer of $\left(\frac{1}{\lambda + \mu_1} + \frac{2\alpha
    \mu_1^2}{\lambda + \mu_1} \right)$ over~$\mu_1 \geq 0.$ The
  statement of the lemma now follows, noting that $\beta^*(\lambda)
  \ra \sqrt[3]{\frac{1}{4\alpha}}$ as $\lambda \da 0.$
\end{proof}

While $\alg_3$ is optimal in light-traffic, it is far from optimal
(not even constant competitive) in heavy traffic. We show this by
designing a certain non-Markovian policy which has $o(\lambda)$ cost
as $\lambda \ra \infty.$

Consider the algorithm~$\alg_3,$ which is defined as follows: Once
number of jobs in the system increases to $U,$ operate at
constant speed~$\mu$ until the system empties. Once system is empty,
set the speed to zero, and remain at this speed until the number of
jobs increases to~$U.$\\ We parameterize $U$ and $\mu$ as
follows:
\begin{align*}
  U &= c_1 \lambda^{\theta_1} \quad (c_1 > 0,\ \theta_1 \leq 1),\\
  \mu &= \lambda + c_2 \lambda^{\theta_2} \quad (c_2 > 0,\ \theta_2 < 1).
\end{align*}
Note that under~$\alg_3,$ the system operates at a single service
rate~$\mu;$ but service commences only once the buffer has built to a
level of $U,$ and continues until the system next empties. A large
value of $U$ (or a small value of~$\mu)$ would clearly result in a
high flow time cost. On the other hand, a small value of~$U$ (or a
large value of~$\mu)$ would result in a high switch cost rate, due to
frequent on-off switching. However, as we show below, an intelligent
intermediate choice results in a cost that scales as $o(\lambda)$ as
$\lambda \ra \infty.$\footnote{We write $f(\lambda) \sim g(\lambda)$
  to mean $\lim_{\lambda \ra \infty} \frac{f(\lambda)}{g(\lambda)} =
  1.$ Similarly, $f(\lambda) \lesssim g(\lambda)$ means $\lim_{\lambda
    \ra \infty} \frac{f(\lambda)}{g(\lambda)} = 0.$}
\begin{theorem}
  Under~$\alg_3,$ as $\lambda \ra \infty,$ $$\displaystyle C_{\alg_3}
  \lesssim c_1 \lambda^{\theta_1} + \frac{1}{c_2} \lambda^{1-\theta_2} +
  2\alpha \frac{c_2}{c_1} \lambda^{1-\theta_1+\theta_2}.$$ Specifically,
  setting $\theta_1 = \frac{2}{3},$ $\theta_2 =
  \frac{1}{3},$ $\displaystyle C_{\alg_3} \lesssim \left(c_1 +
  \frac{1}{c_2} + 2\alpha \frac{c_2}{c_1} \right) \lambda^{\nicefrac{2}{3}}.$
\end{theorem}
Interestingly, under the proposed parameterization of~$\alg_3,$
service is commenced once the buffer level
hits~$O(\lambda^{\nicefrac{2}{3}}),$ at a service rate~$\lambda +
O(\lambda^{\nicefrac{1}{3}}).$ This choice ensures that average buffer
occupancy remains $O(\lambda^{\nicefrac{2}{3}}),$ while the rate of
busy periods scales as $o(1),$ such that the overall switch cost rate
is also $O(\lambda^{\nicefrac{2}{3}}).$ 

\begin{proof}
  We first characterize the switch cost rate under $\alg_3.$ Note that
  the system alternates between busy (server running at speed~$\mu$)
  and idle (server off) periods. Let~$B$ denote a generic busy period,
  and $I$ denote a generic idle period.
  \begin{itemize}
  \item $I$ is the time required to collect $U$ arrivals. Therefore,
    $\Exp{I} = \frac{U}{\lambda} = c_1 \lambda^{\theta_1-1}.$
  \item $B$ is a residual busy period in an M/M/1 queue, with arrival
    rate~$\lambda$ and service rate~$\mu,$ started with the work~$W$
    corresponding to~$U$ jobs. Thus, $$\Exp{B} =
    \frac{\Exp{W}}{1-\nicefrac{\lambda}{\mu}} = \frac{U
      \mu}{\mu-\lambda} \sim \frac{c_1}{c_2}
    \lambda^{1+\theta_1-\theta_2}.$$
  \end{itemize}
Note that since $1 + \theta_1 - \theta_2 > \theta_1 - 1,$ $\Exp{I} =
o(\Exp{B})$ as $\lambda \ra \infty.$ By the renewal reward theorem,
the switch cost rate under~$\alg_3$ is given by $$\frac{2
  \mu^2}{\Exp{B} + \Exp{I}} \sim \frac{2
  \lambda^2}{\nicefrac{c_1}{c_2} \lambda^{1+\theta_1 - \theta_2}} =
\frac{2 c_2}{c_1} \lambda^{1-\theta_1 + \theta_2}.$$

Next, we consider the holding cost rate under~$\alg_3.$ An elementary
sample path argument shows that $$\Exp{N} \leq U + \frac{\lambda}{\mu
  - \lambda} = c_1 \lambda^{\theta_1} +
\frac{\lambda^{1-\theta_2}}{c_2}.$$ This upper bound should be
interpreted as~$U$ plus the mean steady state number in system
corresponding to an M/M/1 queue with arrival rate~$\lambda$ and
service rate~$\mu.$

Combining the above results on the switch cost rate and the holding
cost rate, the statement of the theorem follows.
\end{proof}

\begin{remark}
A \emph{gated static} policy, corresponding to $U = 0,$ does not
attain $o(\lambda)$ cost for \eqref{eq:stochastic_obj_gen}. In contrast, such a policy has been shown to
be constant competitive in a flow time plus (quadratic) speed cost
problem (in the absence of a switching cost);
see~\cite{wierman2009power}.
\end{remark}

To summarize, we find that the stochastic variant of the flow time
plus (quadratic) switching cost problem is significantly easier--a
simple policy is shown to be $(1+2\alpha)$-competitive. Moreover, we
demonstrate that the `multiserver' constraint (i.e., each job can only
be served at a certain fixed rate) implicit in our formulation has a
substantial cost implication; relaxing this constraint results in a
significant cost reduction in heavy traffic.

%% file: SimResultsToBeAdded.tex
\section{Numerical results}\label{sec:sim}
In this section, we present simulation results for the sum of the flow time (per job)+ total switching cost divided by the time horizon. For all simulations, with arrival rate 
$\textsf{arr}$, we generate $\textsf{arr}$ jobs on average per slot and then distribute them over the time horizon arbitrarily. We take the time horizon large, and compute the mean 
flow time (summed across jobs) + total switching cost normalized by the length of the time horizon.

\begin{figure*}
\centering
\begin{tikzpicture}
    \begin{axis}[
        width  = 1*\textwidth,
        height = 7cm,
        major x tick style = transparent,
        ybar,
        bar width=10pt,
        ymajorgrids = true,
        ylabel = {mean flowtime},
        symbolic x coords={$\textsf{arr}=5$, $\textsf{arr}=10$, $\textsf{arr}=15$, $\textsf{arr}=20$},
        xtick = data,
        scaled y ticks = false,
        legend cell align=left,
        legend style={
                at={(.45,.55)},
                anchor=south east,
                column sep=1ex}
    ]

     \addplot[style={bblue,fill=black,mark=none}]
            coordinates {($\textsf{arr}=5$, 11.751) ($\textsf{arr}=10$,22.67) ($\textsf{arr}=15$,33.39) ($\textsf{arr}=20$, 44.12)};

        \addplot[style={bblue,fill=bblue,mark=none}]
            coordinates {($\textsf{arr}=5$, 9.89) ($\textsf{arr}=10$,17.08) ($\textsf{arr}=15$,23.64) ($\textsf{arr}=20$,30.13)};

        \addplot[style={bblue,fill=rred,mark=none}]
            coordinates {($\textsf{arr}=5$, 9.3) ($\textsf{arr}=10$,15.78) ($\textsf{arr}=15$, 21.6) ($\textsf{arr}=20$,27.36)};
            
        \legend{\text{Balance} $s(t) = n(t) /\alpha$,  \text{Balance} \ $|s(t)-s(t-1)| = n(t)$, Proposed Algorithm $n(t)/\alpha^{1/4}$}
    \end{axis}
\end{tikzpicture}
\caption{Comparison of mean flow time with different algorithms as a function of mean arrival rate per slot with linear switching cost with $\alpha=2$.}
\label{fig:linscalpha2} 
\end{figure*}
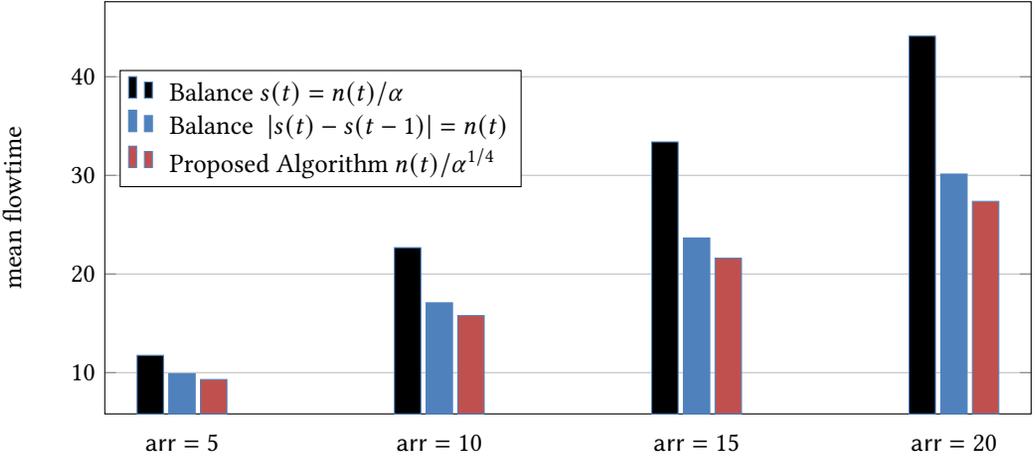

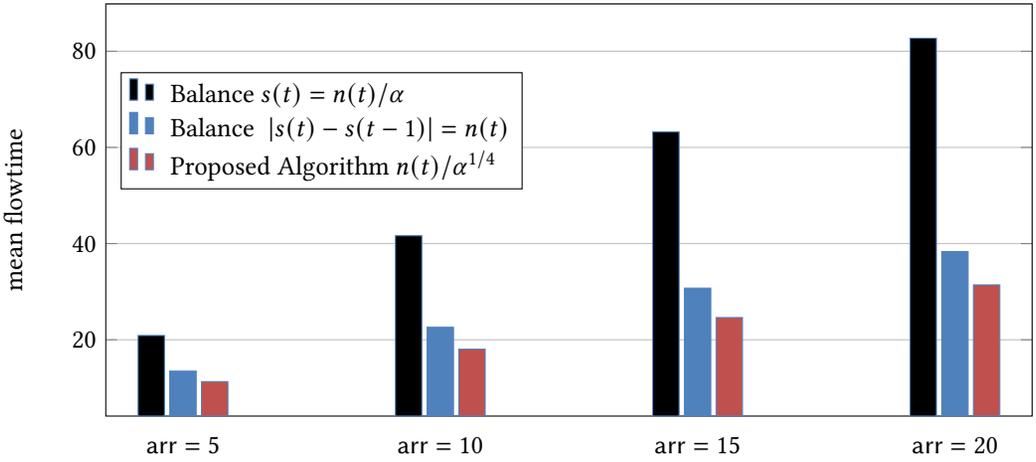
\begin{figure*}
\centering
\begin{tikzpicture}
    \begin{axis}[
        width  = 1*\textwidth,
        height = 7cm,
        major x tick style = transparent,
        ybar,
        bar width=10pt,
        ymajorgrids = true,
        ylabel = {mean flowtime},
        symbolic x coords={$\textsf{arr}=5$, $\textsf{arr}=10$, $\textsf{arr}=15$, $\textsf{arr}=20$},
        xtick = data,
        scaled y ticks = false,
        legend cell align=left,
        legend style={
                at={(.45,.55)},
                anchor=south east,
                column sep=1ex}
    ]

     \addplot[style={bblue,fill=black,mark=none}]
            coordinates {($\textsf{arr}=5$, 20.89) ($\textsf{arr}=10$,41.64) ($\textsf{arr}=15$,63.22) ($\textsf{arr}=20$, 82.7)};

        \addplot[style={bblue,fill=bblue,mark=none}]
            coordinates {($\textsf{arr}=5$, 13.51) ($\textsf{arr}=10$,22.6) ($\textsf{arr}=15$,30.72) ($\textsf{arr}=20$,38.35)};

        \addplot[style={bblue,fill=rred,mark=none}]
            coordinates {($\textsf{arr}=5$, 11.28) ($\textsf{arr}=10$,18.05) ($\textsf{arr}=15$, 24.63) ($\textsf{arr}=20$,31.39)};
            
        \legend{\text{Balance} $s(t) = n(t) /\alpha$,  \text{Balance} \ $|s(t)-s(t-1)| = n(t)$, Proposed Algorithm $n(t)/\alpha^{1/4}$}
    \end{axis}
\end{tikzpicture}
\caption{Comparison of mean flow time with different algorithms as a function of mean arrival rate per slot with linear switching cost with $\alpha=4$.}
\label{fig:linscalpha4} 
\end{figure*}

We start with linear switching cost case with $\alpha=1$ and $\alpha=2$ for arrival rates $5,10,15,20$. For $\alpha=1$, in Fig. \ref{fig:linearscalpha1}, we compare the performance of the proposed algorithm 
$s(t)=n(t)$ with  an algorithm that chooses $s(t)=n(t)/2$ and the {\it Balance} algorithm $|s(t)-s(t-1)| = n(t)$ . 
Next, we consider $\alpha=2$ and $\alpha=4$, 
in Fig. \ref{fig:linscalpha2} and Fig. \ref{fig:linscalpha4}, respectively, and compare the performance of the proposed algorithm with the {\it Balance} algorithms. Fig. \ref{fig:linearscalpha1}, Fig. \ref{fig:linscalpha2} and Fig. \ref{fig:linscalpha4} reveal that the performance of the 
proposed algorithm is better than other algorithms.

The quadratic switching case is considered next, where we consider the case of $\alpha=1, 2, 4$ for arrival rates $5,10,15,20$ for different choice of parameter $\beta$ of the proposed algorithm and {\it Balance} algorithm in Figs. \ref{fig:quadscalpha1}, and Fig. \ref{fig:quadscalpha2}. These numerical results are interesting, where at times a larger arrival rate leads to smaller sum of the flow time (per job)+ total switching cost divided by the time horizon, since when there are larger number of jobs, the number of active servers is chosen large early enough and does not need to changed often enough over the time horizon, leading to small switching cost together with small flow time.

\begin{figure*}
\centering
\begin{tikzpicture}
    \begin{axis}[
        width  = 1*\textwidth,
        height = 7cm,
        major x tick style = transparent,
        ybar,
        bar width=10pt,
        ymajorgrids = true,
        ylabel = {mean flowtime},
        symbolic x coords={$\textsf{arr}=1000$},
        xtick = data,
        scaled y ticks = false,
        legend cell align=left,
        legend style={
                at={(1,.55)},
                anchor=south east,
                column sep=1ex}
    ]

     \addplot[style={bblue,fill=black,mark=none}]
            coordinates {($\textsf{arr}=1000$,153710) };

        \addplot[style={bblue,fill=bblue,mark=none}]
            coordinates {($\textsf{arr}=1000$, 130590) };

        \addplot[style={bblue,fill=rred,mark=none}]
            coordinates {($\textsf{arr}=1000$, 58463)};
            
        \addplot[style={bblue,fill=ggreen,mark=none}]
            coordinates {($\textsf{arr}=1000$, 35549)};

        \legend{ $\beta=\sqrt{3}$, $\beta=2$, $\beta=4$, Balance $\alpha(s(t)-s(t-1))^2 = n(t)$}
    \end{axis}
\end{tikzpicture}
\caption{Comparison of mean flow time with the proposed algorithm with different values of $\beta$ and Balance $\alpha (s(t)-s(t-1))^2 =  n(t)$ for quadratic switching cost with 
$\alpha=2$ for extremely large mean arrival
 rate of $1000$ per slot.}
\label{fig:quadscalpha2extremearrrate} 
\end{figure*}
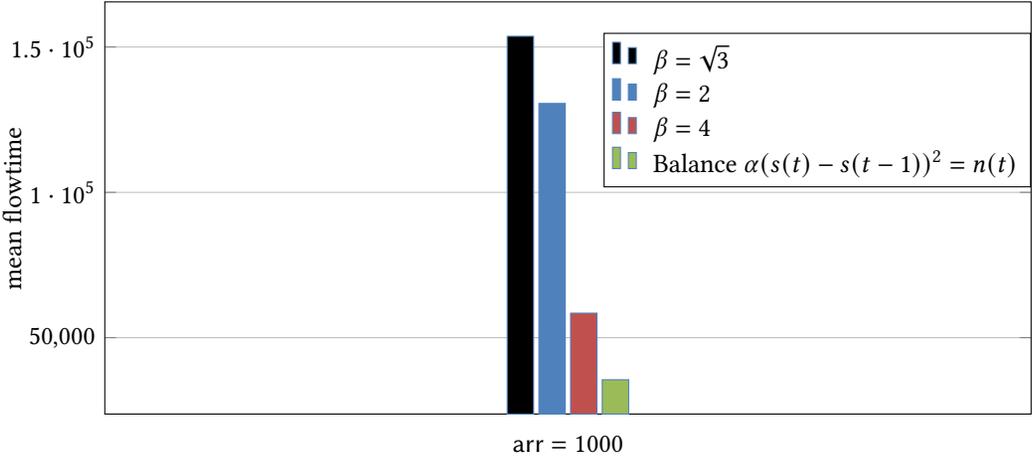

From Figs. \ref{fig:quadscalpha1}, and  \ref{fig:quadscalpha2}, it also appears that the 
performance of the proposed algorithm improves as $\beta$ is increased from $1$ to $4$, and the performance with $\beta=4$ is similar to that of the  {\it Balance} algorithm. Recall that in our theoretical guarantee, the optimal choice of $\beta$ was close to $2$. From Fig. \ref{fig:quadscalpha1} and Fig. \ref{fig:quadscalpha2} 
it appears that as the arrival rate increases the performance of the proposed algorithm degrades compared to the {\it Balance} algorithm. To numerically verify that the performance of the proposed algorithm with $\beta=2$ is not unboundedly worse compared to the the {\it Balance} algorithm, in Fig. \ref{fig:quadscalpha2extremearrrate} we consider significantly large arrival rate of $1000$ and show that the ratio of the flow time of the proposed algorithm with $\beta=2$  and the balance algorithm is less than $2$.


%

\begin{figure*}
\centering
\begin{tikzpicture}
    \begin{axis}[
        width  = 1*\textwidth,
        height = 7cm,
        major x tick style = transparent,
        ybar,
        bar width=10pt,
        ymajorgrids = true,
        ylabel = {mean flowtime},
        symbolic x coords={$\textsf{arr}=5$, $\textsf{arr}=10$, $\textsf{arr}=15$, $\textsf{arr}=20$},
        xtick = data,
        scaled y ticks = false,
        legend cell align=left,
        legend style={
                at={(.55,.65)},
                anchor=south east,
                column sep=1ex}
    ]

     \addplot[style={bblue,fill=black,mark=none}]
            coordinates {($\textsf{arr}=5$, 7.50) ($\textsf{arr}=10$,13.5661) ($\textsf{arr}=15$,19.43) ($\textsf{arr}=20$, 25.21)};

        \addplot[style={bblue,fill=bblue,mark=none}]
            coordinates {($\textsf{arr}=5$, 7.57) ($\textsf{arr}=10$,13.7) ($\textsf{arr}=15$,19.56) ($\textsf{arr}=20$,25.16)};

        \addplot[style={bblue,fill=rred,mark=none}]
            coordinates {($\textsf{arr}=5$, 10.6) ($\textsf{arr}=10$,20.98) ($\textsf{arr}=15$, 31.6) ($\textsf{arr}=20$,41.84)};
            
        \legend{$s(t) = n(t)$,  \text{Balance} \ $|s(t)-s(t-1)| = n(t)$, $s(t) = n(t)/2$}
    \end{axis}
\end{tikzpicture}
\caption{Comparison of mean flow time with different algorithms as a function of mean arrival rate per slot with linear switching cost with $\alpha=1$	.}
\label{fig:linearscalpha1} 
\end{figure*}
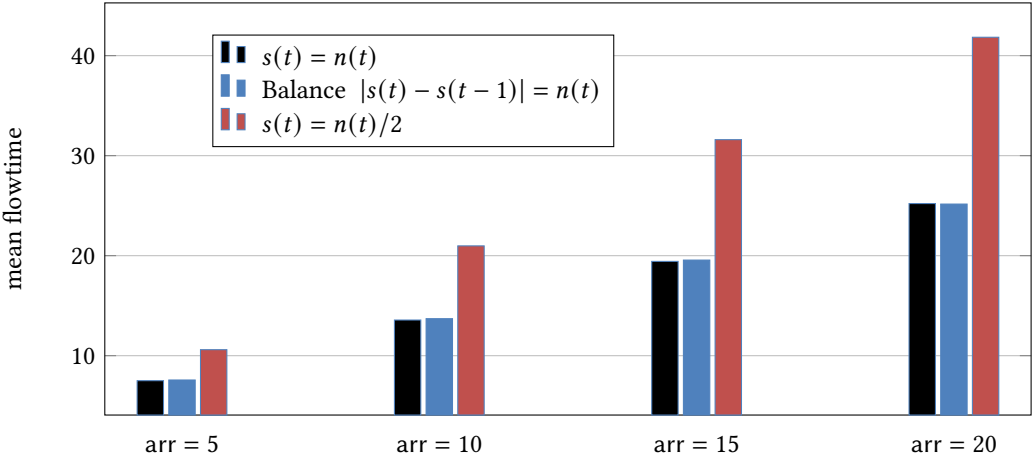

\begin{figure*}
\centering
\begin{tikzpicture}
    \begin{axis}[
        width  = 1*\textwidth,
        height = 7cm,
        major x tick style = transparent,
        ybar,
        bar width=10pt,
        ymajorgrids = true,
        ylabel = {mean flowtime},
        symbolic x coords={$\textsf{arr}=5$, $\textsf{arr}=10$, $\textsf{arr}=15$, $\textsf{arr}=20$},
        xtick = data,
        scaled y ticks = false,
        legend cell align=left,
        legend style={
                at={(.55,.55)},
                anchor=south east,
                column sep=1ex}
    ]

     \addplot[style={bblue,fill=black,mark=none}]
            coordinates {($\textsf{arr}=5$, 21.55) ($\textsf{arr}=10$,91.99) ($\textsf{arr}=15$,210.62) ($\textsf{arr}=20$, 376.63)};

        \addplot[style={bblue,fill=bblue,mark=none}]
            coordinates {($\textsf{arr}=5$, 8.64) ($\textsf{arr}=10$,31.09) ($\textsf{arr}=15$,70.9) ($\textsf{arr}=20$,127.3831)};

        \addplot[style={bblue,fill=rred,mark=none}]
            coordinates {($\textsf{arr}=5$, 8.87) ($\textsf{arr}=10$,24.63) ($\textsf{arr}=15$, 53.83) ($\textsf{arr}=20$,96.53)};
            
        \addplot[style={bblue,fill=ggreen,mark=none}]
            coordinates {($\textsf{arr}=5$, 11.28) ($\textsf{arr}=10$,29.18) ($\textsf{arr}=15$, 32.6) ($\textsf{arr}=20$,30.08)};
            
             \addplot[style={bblue,fill=ppurple,mark=none}]
            coordinates {($\textsf{arr}=5$, 12.08) ($\textsf{arr}=10$,23.55) ($\textsf{arr}=15$, 35.34) ($\textsf{arr}=20$,46.69)};
            
        \legend{$\beta=1$, $\beta=\sqrt{3}$, $\beta=2$, $\beta=4$, Balance $\alpha(s(t)-s(t-1))^2 = n(t)$}
    \end{axis}
\end{tikzpicture}
\caption{Comparison of mean flow time with the proposed algorithm with different values of $\beta$ and Balance $\alpha (s(t)-s(t-1))^2 =  n(t)$ for quadratic switching cost with $\alpha=1$ as a function of mean arrival
 rate per slot.}
\label{fig:quadscalpha1} 
\end{figure*}
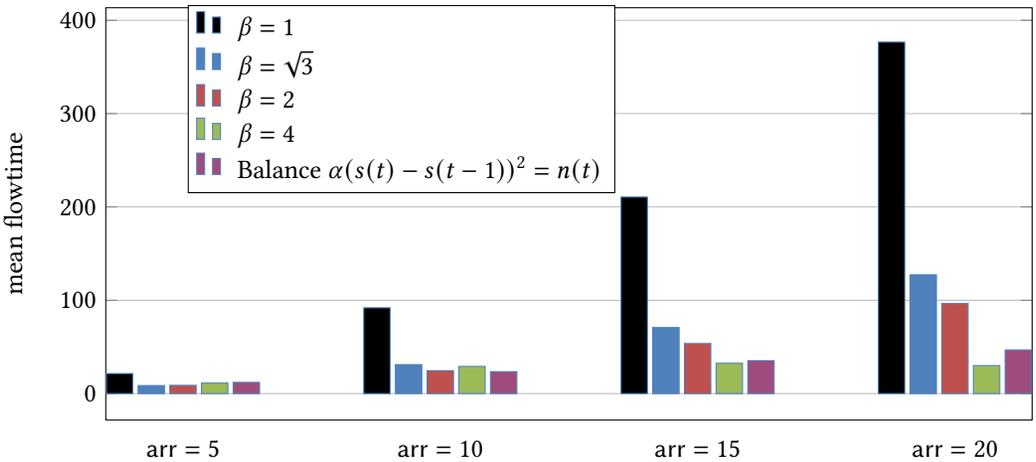

\begin{figure*}
\centering
\begin{tikzpicture}
    \begin{axis}[
        width  = 1*\textwidth,
        height = 7cm,
        major x tick style = transparent,
        ybar,
        bar width=10pt,
        ymajorgrids = true,
        ylabel = {mean flowtime},
        symbolic x coords={$\textsf{arr}=5$, $\textsf{arr}=10$, $\textsf{arr}=15$, $\textsf{arr}=20$},
        xtick = data,
        scaled y ticks = false,
        legend cell align=left,
        legend style={
                at={(.5,.55)},
                anchor=south east,
                column sep=1ex}
    ]

     \addplot[style={bblue,fill=black,mark=none}]
            coordinates {($\textsf{arr}=5$,41.04) ($\textsf{arr}=10$,178.33) ($\textsf{arr}=15$,411.16) ($\textsf{arr}=20$, 732.84)};

        \addplot[style={bblue,fill=bblue,mark=none}]
            coordinates {($\textsf{arr}=5$, 14.77) ($\textsf{arr}=10$,63.06) ($\textsf{arr}=15$,142) ($\textsf{arr}=20$,252)};

        \addplot[style={bblue,fill=rred,mark=none}]
            coordinates {($\textsf{arr}=5$, 11.5) ($\textsf{arr}=10$,46.7) ($\textsf{arr}=15$, 106) ($\textsf{arr}=20$,192)};
            
        \addplot[style={bblue,fill=ggreen,mark=none}]
            coordinates {($\textsf{arr}=5$, 23.13) ($\textsf{arr}=10$,20.49) ($\textsf{arr}=15$, 30.6) ($\textsf{arr}=20$,52)};
            
             \addplot[style={bblue,fill=ppurple,mark=none}]
            coordinates {($\textsf{arr}=5$, 16.75) ($\textsf{arr}=10$,32.3) ($\textsf{arr}=15$, 48.47) ($\textsf{arr}=20$,63.07)};
            
        \legend{$\beta=1$, $\beta=\sqrt{3}$, $\beta=2$, $\beta=4$, Balance $\alpha(s(t)-s(t-1))^2 = n(t)$}
    \end{axis}
\end{tikzpicture}
\caption{Comparison of mean flow time with the proposed algorithm with different values of $\beta$ and Balance $\alpha (s(t)-s(t-1))^2 =  n(t)$ for quadratic switching cost with 
$\alpha=2$ as a function of mean arrival
 rate per slot.}
\label{fig:quadscalpha2} 
\end{figure*}
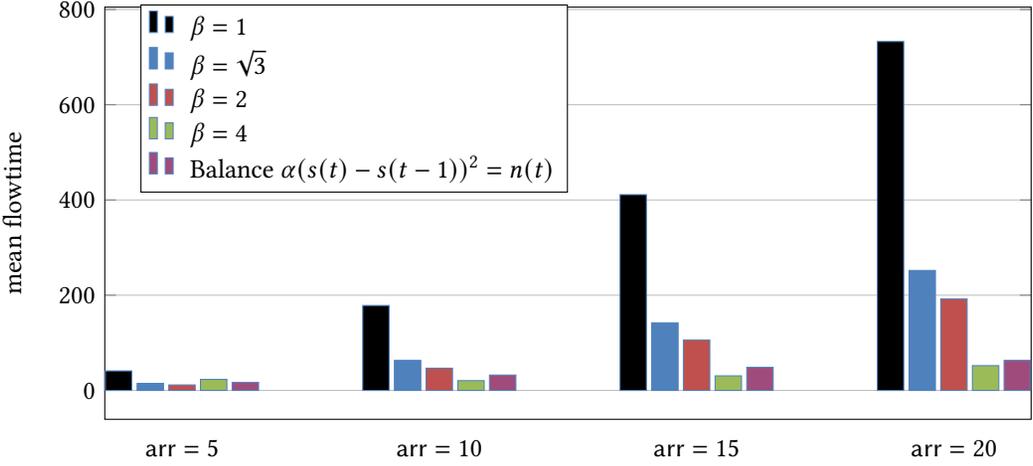

%% file: Conclusions.tex
\section{Conclusions}
In this paper, we have broadened the horizon of OCO-S problem by
incorporating two important aspects: i) the objective function has unbounded memory, i.e., at any
time depends on all the actions taken so far, and ii) the considered objective
function motivated by capacity provisioning where the queuing dynamics interact with server
switching cost  is non-convex. Prior work on OCO indicated that the regret or the competitive ratio 
increases linearly in the memory size, however, using the specific 
structure of the considered problem, we avoided the negative results, and 
derived algorithms with memory independent competitive ratios.

%% file: MissingProofs.tex
\section{appendices}
\section{Missing Proofs}
\begin{proof}[Proof of Lemma \ref{lem:alphale1}]
 With $\cA_f$ each outstanding job is under service from one server at any point of time, which is the best possible from any job's point of view in minimizing its flow time. 
Thus, the flow time of any job is just $\sum_{t} n(t)$, where $n(t)$ is the number of jobs present in the system when each job is processed at unit speed as soon as it arrives. 
Moreover, the 
total switching cost  of $\cA$ at time $t$ is at most $\alpha n(t)$. Therefore, the total cost of $\cA$ is at most $(1+\alpha)  \sum_{t} n(t)$. 
From the above discussion we get that $\opt$'s flow time $ \sum_{t} n_o(t) \ge \sum_{t} n(t)$, since  any job can be processed by at most one with unit speed at any time. Thus, $(1+\alpha)  \sum_{t} n(t) \le (1+\alpha)  \sum_{t} n_o(t)$, and  we get that the competitive ratio of $\cA$ is at most $1+\alpha$. Given $\alpha\le 1$, the competitive ratio is at most $2$.

We next provide an instance where the competitive ratio of $\cA_f$ is close to $2$.
Consider an input where at times $ 2, 4, \dots, 2k$ for some $k\in\bbZ^+$, $x$ (even) number of jobs arrive, each with size $1$. The proposed algorithm will choose $s(t) =x$ for time slots $ 2, 4, \dots, 2k$ and $s(t)=0$ otherwise, resulting in a total flow time of $x\cdot k$ and a total switching cost of $2kx$. In contrast, consider another algorithm $\cB$ that processes $x/2$ jobs in each time slot starting from time slot $2$. $\cB$ pays a 
switching cost of $x/2$ only at time $2$ and at time $2k+1$ making the total switching cost just $x$. Moreover, the flow time of  
algorithm $\cB$ is $k(x/2+2x/2) = 1.5kx$. Since $\opt$ is at least as good as $\cB$, the competitive ratio of the proposed algorithm is at least $\frac{3xk}{1.5xk+x} \approx 2$ choosing $k$ large.
\end{proof}

\begin{proof} [Proof of Lemma \ref{lem:balanced}]
  
We begin with $\cA_{bal-1}$
Let the total number of jobs be $\sfN$ and all the $\sfN$ jobs arrive at time $0$ and let $\alpha=\sfN$. Then the algorithm $\cA_{bal-1}$ keeps $s(t)=1$ always,  and its flow time is $\Omega(\sfN^2)$. An alternate algorithm $\cB$ that 
uses $s(t) = \frac{\sfN}{\sqrt{\alpha}}$ at time $0$ and until all jobs are finished has a switching cost of $2 \sqrt{\alpha} \sfN = O(\sfN^{3/2})$ and a flow time of $\sqrt{\alpha} \sfN=O(\sfN^{3/2})$. Since $\opt$ is at least as good as $\cB$, we get that the competitive ratio of $\cA_{bal-1}$ is $\Omega(\sfN^{1/2})$. Choosing $\sfN$ large we get the result. 

Next, we consider the algorithm $\cA_{bal-2}$ with the input where $\sfN$ jobs arrive at time $0$ and $\alpha=\sfN^2$. Algorithm $\cA_{bal-2}$ has a switching cost of $\Omega(\sfN^{2.5})$, while an alternate algorithm $\cB$ that uses $s(t) = \frac{\sfN}{\sqrt{\alpha}}$ at time $0$ and until all jobs are finished has a switching cost of 
$2 \sqrt{\alpha} \sfN = O(\sfN^{2})$ and a flow time of $\sqrt{\alpha} \sfN=O(\sfN^{2})$. Since $\opt$ is at least as good as $\cB$, we get that the competitive ratio of $\cA_{bal-1}$ is $\Omega(\sfN^{1/2})$. Choosing $\sfN$ large we get the result. 
\end{proof}

\begin{proof}[Proof of Lemma \ref{lem:linearopt}] In light of linear switching cost and all jobs being available at time $0$, an optimal algorithm will choose the maximum speed $s_{\max}$ at time $1$ itself. Moreover, it is also optimal to maintain the maximum speed $s_{\max}$ as long as possible without violating the work neutrality constraint, and thereafter to choose $s(t) = n(t)$. With $\sfN$ as the total number of jobs,  the problem boils down to finding $s_{\max}$ that solves 
$$\min_{s_{\max}\ge 0} \quad \left(\sum_{t=1}^{T_{\max}} t s_{\max} \right)+  2 s_{\max},$$
where $T_{\max}$ is the largest time for which maintaining a constant speed $s_{\max}$ does not violate the work neutrality constraint. Excluding the minor edge effects at the end, choosing $s_{\max} = \frac{\sqrt{\sfN(\sfN-1)}}{2\alpha^{1/2}}$ is optimal.
\end{proof}

\begin{proof} [Proof of Lemma \ref{lbcrlinearcase}]
Consider two input sequences $\sigma_1$ and $\sigma_2$. With $\sigma_1$, $\sfN$ jobs arrive at time $1$ and no more jobs arrive thereafter, while with $\sigma_2$, 
$\sfN$ jobs arrive at each time $t=1, \dots, T$.

Let $\cA_\gamma$ be an online algorithm that chooses $s(t) = \frac{n(t)}{\alpha^\gamma}$ for $\gamma\ge 0$. If $\gamma < 1/4$ then the input is $\sigma_1$, else if $\gamma \ge \frac{1}{4} $ then the input is $\sigma_2$,.

With $\sigma_1$, the $\opt$ chooses $s(t) = \frac{n(t)}{\sqrt{\alpha}}$ (Lemma \ref{lem:linearopt}), and the total cost of $\opt$, $C_\opt(\sigma_1) \le 3\sqrt{\alpha}\sfN$.  
$\cA_\gamma$ on the other hand has a switching cost of $\alpha^{1-\gamma} \sfN$ and thus, 
\begin{equation}\label{eq:crlb1}
\frac{C_{\cA_\gamma}(\sigma_1)}{C_\opt(\sigma_1)}\ge \Omega( \alpha^{\frac{1}{2}-\gamma})
\end{equation} for $\gamma < 1/4$.

Next, consider $\sigma_2$, where $\sfN$ jobs arrive at each time slot for $t=1, \dots, T$, we will pick $T$ large later. 
Consider an algorithm $\cB$ that chooses $s(1) = \sfN$ and $s(t)=s(1)$ until time $T$. The cost of $\cB$ is at most 
$\alpha \sfN + \sfN T$. Given that $\opt$ is at least as good as  $\cB$, $C_\opt(\sigma_2) \le \alpha \sfN + \sfN T$.

For $\cA_{\gamma}$, that chooses $s(t) = \frac{n(t)}{\alpha^\gamma}$, the number of outstanding jobs at time $n(t)$ satisfies the recursion $n(t) = n(t-1) + \sfN  - \frac{n(t-1) + \sfN}{\alpha^\gamma}$.
Simple algebra\footnote{
Fixed point of recursion $n(t) = n(t-1) + \sfN  - \frac{n(t-1) + \sfN}{\alpha^\gamma}$ is 
$n(t) = (\alpha^\gamma-1) \sfN$.} reveals that for any $t$, $\tau < t \le T$ where $\tau$ is $o(T)$,   the number of jobs remaining $n(t) = (\alpha^\gamma-1) \sfN$. 
Thus, the flow time of $\cA_\gamma$ is at least $\sfN(\alpha^\gamma-1) T$. Thus, $C_{\cA_{\gamma}} \ge  \Omega( \sfN \alpha^\gamma T)$. Choosing $T$ large enough compared to $\alpha$, for large enough $\alpha$, we get that 
\begin{equation}\label{eq:crlb2}\frac{C_{\cA_{\gamma}}(\sigma_2)}{C_{\opt}(\sigma_2)} \ge \Omega(\alpha^\gamma).
\end{equation}

Combining \eqref{eq:crlb1} and \eqref{eq:crlb2}, we get that  
 the competitive ratio of $\cA_\gamma$ is $\ge \max\{\alpha^{\gamma},\alpha^{\frac{1}{2}-\gamma}\}$. Thus, to minimize its competitive ratio $\cA_\gamma$ should choose $\gamma=\frac{1}{4}$, and the resulting competitive ratio lower bound is $\Omega(\alpha^{1/4})$.
\end{proof}

\begin{proof}[Proof of Lemma \ref{eq:costalg}]
Consider any time $t$. One part of cost \eqref{eq:probquad} is simply $n(t)$. All we need to do is to upper bound the switching cost of $\textsf{alg}$ as a function of 
$n(t)$. 
At time $t$, either the number of active servers increases, decreases or remains the same. 

If it increases, then the switching cost paid  at time $t$ is 
at most $\beta^2 n(t)$. Moreover, corresponding to this increase,  the number of active servers has to decrease either gradually or suddenly over time, since  the number of active servers 
has to become zero whenever all the remaining work is complete (work neutrality constraint). Thus, corresponding to the increase in the number of active servers at time $t$, in the worst case the number of active servers becomes zero suddenly at some future time in which case the switching (down) cost is at most $\beta^2 n(t)$. If the decrease is gradual, then the switching cost will only be smaller, since the switching cost is quadratic in the difference between the number of active servers in consecutive time.
Thus, the total switching cost (counting both the switching up and switching down cost across time) is at most $2 \beta^2 \sum_t n(t)$. 
\end{proof} 

%% file: CorrectDualLowerBound
\subsubsection{Proof of Theorem \ref{thm:main}}

To prove Theorem \ref{thm:main}, we will use the primal-dual scheme. 
Towards that end, we first rewrite a relaxed version of  problem \eqref{eq:probquad} by writing the total flow time as the sum of flow time of each job  as follows

\begin{equation}\label{eq:probeq}
\begin{aligned}
& \underset{s(t)}{\min}
& & \sum_j \left(\sum_{t=a_j}^{d_j} s_j(t) \right) \frac{d_j-a_j}{w_j}+ \alpha \sum_{t}  (s(t)- s(t-1))^2 \\
& \text{}
& & \begin{aligned} \sum_{t=a_j}^{d_j} s_j(t)  & \ge w_j, \ \forall \ j, \\
 s_j(t) &\geq 0 \ \forall \ t, j, \\
 \sum_{\tau \le t} s(\tau) &\le \sum_{\tau\le t} w(\tau) \ \forall \ t,\\
  \sum_{j} s_j(t) &= s(t) \ \forall \ t, 
 \end{aligned}
\end{aligned}
\end{equation}
where $s_j(t)=1$ if  job $j$ is being processed at time $t$ and $0$ otherwise, and $s(t)$ is the total number of active servers at time $t$, and the constraint $\sum_{t=a_j}^{d_j} s_j(t) \ge w_j$ implies that job $j$ is eventually completed.
The constraint $\sum_{\tau \le t} s(\tau) \le \sum_{\tau\le t} w(\tau) \ \forall \ t,$ corresponds to the work neutrality constraint and for brevity we denote it as 
$\mathsf{acc}(t)\le D(t)$ hereafter, where $\mathsf{acc}(t)=\sum_{\tau \le t} s(\tau)$, while $D(t)= \sum_{\tau\le t} w(\tau)$.

The Lagrangian dual function associated with \eqref{eq:probeq} is \newline $
 \min_{s(t), \ d_j, \ \mathsf{acc}(t)\le D(t) } \cL(\lambda_j, \mu_j, \nu_t)=$
\begin{align*}\label{} &  \min_{s(t), \ d_j, \ \mathsf{acc}(t)\le D(t)} \quad  \sum_j \left(\sum_{t=a_j}^{d_j} s_j(t)  \right) \frac{d_j-a_j}{w_j} + \alpha \sum_{t}  (s(t)- s(t-1))^2 \\ 
& \quad \quad + \sum_j \lambda_j \left(w_j - \sum_{t=a_j}^{d_j} s_j(t)\right)   + \sum_j \mu_j s_j(t) + \sum_t \nu_t (\sum_{j} s_j(t) - s(t)),
\end{align*}
where  $\lambda_j, \mu_j,$ and $\nu_t$ are the dual variables corresponding to $\sum_{t=a_j}^{d_j} s_j(t)   \ge w_j$, $s_j(t) \geq 0 \ \forall \ t, j$, and $\sum_{j} s_j(t) = s(t)$, respectively, and {\bf we have kept the work neutrality constraint as is}. Keeping the work neutrality constraint as it is is in fact critical for the analysis. Instead if it is included inside the Lagrangian, that results in a trivial lower bound that is not useful. This feature makes the analysis sufficiently novel compared to usual application of primal-dual scheme for similar problems.
Note that $$   \sum_j \left(\sum_{t=a_j}^{d_j} s_j(t)  \right) \frac{d_j-a_j}{w_j}  + \alpha \sum_{t}  (s(t)- s(t-1))^2 + \sum_j \lambda_j \left(w_j - \sum_{t=a_j}^{d_j} s_j(t)\right)$$ is also equal to 
\begin{align}\label{eq:dummy1}
\sum_j \lambda_j w_j -   \sum_{j}  \sum_{t=a_j}^{d_j} s_j(t)\left(\lambda_j -  \frac{d_j-a_j}{w_j}\right)  +\alpha \sum_{t}  (s(t)- s(t-1))^2.
\end{align}

Recall that $\max_{\lambda_j, \mu_j, \nu_t} \min_{s(t), \ d_j, \ \mathsf{acc}(t)\le D(t)} \cL(\lambda_j, \mu_j, \nu_t)$ is a lower bound on the objective value of the primal problem \eqref{eq:probeq}. 
Thus, even if we consider a fixed value of  $\lambda_j, \mu_j, \nu_t$, $$\min_{s(t), \ d_j, \ \mathsf{acc}(t)\le D(t)} \cL(\lambda_j, \mu_j, \nu_t)$$ is a lower bound on the objective value of the primal problem. In particular, we will make the following choices, $\mu_j=\nu_t=0\ \forall \ j,t$. How to choose $\lambda_j$ will be defined next.

Using \eqref{eq:dummy1}, and setting dual variables $\mu_j=\nu_t=0, \eta_t=0 \ \forall \ j, t$,  we get 
$$\min_{s(t), \ d_j, \ \mathsf{acc}(t)\le D(t)} \cL (\lambda_j, \mu_j=0, \nu_t=0) = $$
\begin{align}\nn & \sum_j \lambda_j w_j -  \max_{s(t), \ d_j, \ \mathsf{acc}(t)\le D(t)} \left(\sum_{j}  \sum_{t=a_j}^{d_j} s_j(t)\left(\lambda_j -  \frac{d_j-a_j}{w_j}\right) \right. \\ \label{eq:lbdual1}
& \quad \left.-\alpha \sum_{t}  (s(t)- s(t-1))^2\right).
\end{align}

The fact that for job $j$ the sum over time in the second term is for indices $t\le d_j$, \eqref{eq:lbdual1} is also equal to
\begin{align}\nn & \sum_j \lambda_j w_j -  \max_{s(t), \ d_j, \ \mathsf{acc}(t)\le D(t)} \left(\sum_{j}  \sum_{t=a_j}^{d_j} s_j(t)\left(\lambda_j -  \frac{t-a_j}{w_j}\right) \right. \\ \label{eq:lbdual2}
& \quad \left. -\alpha \sum_{t}  (s(t)- s(t-1))^2\right).
\end{align}

{\bf Choosing the dual variable $\lambda_j$} Given the $\textsf{alg}$, choose $\lambda_j$ such that
$\lambda_j w_j = \Delta_j F_{\textsf{alg}}$, where $\Delta_j F_{\textsf{alg}}$ is the increase in the total flow time of the $\textsf{alg}$ because of arrival of job $j$ with size $w_j$ at time $a_j$, disregarding jobs arriving after time $a_j$.

For this choice of $\lambda_j$, we obtain the following important structural result for $\textsf{alg}$. 
\begin{lemma}\label{lem:lambdajtime}
\begin{equation}\label{eq:lambdajtconnection}
\lambda_j - \left(\frac{t-a_j}{w_j}\right) \le \frac{3}{\beta} (\alpha n_{\textsf{alg}}(t))^{1/2}
\end{equation} for all $t$ and $\beta \ge 1$, where 
$n_{\textsf{alg}}(t)$ is the number of remaining jobs with  $\textsf{alg}$ at time $t$. 
\end{lemma}
Using \eqref{eq:lambdajtconnection}, we next lower bound the dual cost \eqref{eq:lbdual2}. In particular, we upper bound 
 \begin{equation}\label{eq:dummy}
 \max_{s(t), \ d_j, \ \mathsf{acc}(t)\le D(t)} \left(\sum_{j}  \sum_{t=a_j}^{d_j} s_j(t)\left(\frac{3}{\beta} (\alpha n_{\textsf{alg}}(t))^{1/2}\right)  -\alpha \sum_{t}  (s(t)- s(t-1))^2\right).
\end{equation}
Rewriting, \eqref{eq:dummy} is also equal to 
 \begin{equation}\label{eq:objinterest1}
\max_{s(t), \ d_j, \ \mathsf{acc}(t)\le D(t)} \sum_{t} s(t)\left(\frac{3}{\beta} (\alpha n_{\textsf{alg}}(t))^{1/2} - \alpha s(t)\right) + \alpha \sum_{t}  \left(s(t-1)^2 - 2 s(t)s(t-1))\right).
 \end{equation}
 
 Note that $s(t)$, the variable to be optimized in \eqref{eq:objinterest1}, is not the choice made by $\textsf{alg}$ but by any algorithm subject to the constraints. To avoid this confusion, we rewrite \eqref{eq:objinterest1} as 
  \begin{equation}\label{eq:objinterest}
\max_{x(t), \ d_j, \ \mathsf{acc}(t)\le D(t)} \sum_{t} x(t)\left(\frac{3}{\beta} (\alpha n_{\textsf{alg}}(t))^{1/2} - \alpha x(t)\right) + \alpha \sum_{t}  \left(x(t-1)^2 - 2 x(t)x(t-1))\right),
 \end{equation}
 by replacing $s(t)$ by $x(t)$, and the constraints are over $x(t)$ now.
 
 Define a busy cycle $\sfC_i$ for $\textsf{alg}$ as an interval  of time $T_i = [t_i, t_{i+1}]$ where $n_{\textsf{alg}}(t)>0$ for all $t\in T_i$, and let the total
 number of job arrivals in $T_i$ be $\sfA_i$. Clearly, by the very definition, $\textsf{alg}$ completes all the jobs $\sfA_i$ that arrive in  $\sfC_i$ during $T_i$. 
 
From \eqref{eq:objinterest}, it is clear that to maximize  \eqref{eq:objinterest}, $x(t)$ will be non-zero only when $n_{\textsf{alg}}(t) > 0$, i.e. only during some busy cycle of  $\textsf{alg}$. This follows since otherwise, the contribution for $t$ such that $n_{\textsf{alg}}(t) = 0$ is $-\alpha \sum_{t}  (x(t)- x(t-1))^2$ which is non-positive.
Next consider a busy cycle $\sfC_i$ for $\textsf{alg}$. Because of the work neutrality constraint for each algorithm, $\sum_{t\in T_i}x(t) \le \sum_{j\le i} \sfA_j$, all the jobs that have arrived by the end of cycle $\sfC_i$. One choice to maximize  \eqref{eq:objinterest}, is to keep $\sum_{t\in T_i}x(t) = \sfA_i$ (competely exhaust the limit in each busy cycle of $\textsf{alg}$) or to accumulate jobs and assign them to a subset of cycles of $\textsf{alg}$, where possibly larger $n_{\textsf{alg}}(t)$ exist in future. Lets consider the case when $\sum_{t\in T_i}x(t) = \sfA_i$ first. 
In this case, for any cycle $C_i$ \begin{equation}\label{eq:objinterest1}
\max_{x(t), \sum_{t\in T_i}x(t) = \sfA_i} \sum_{t \in T_i} x(t)\left(\frac{3}{\beta} (\alpha n_{\textsf{alg}}(t))^{1/2} - \alpha x(t)\right) + \alpha \sum_{t\in T_i}  \left(x(t-1)^2 - 2 x(t)x(t-1))\right).
\end{equation} 
We upper bound the two terms separately. 
 Clearly, even without putting any constraints on $x(t)$, $$x(t)\left(\frac{3}{\beta}\sqrt{\alpha n_{\textsf{alg}}(t)} - \alpha x(t)\right)$$ is maximized at  
 $x(t) = \frac{\sqrt{n_{\textsf{alg}}(t)}}{\frac{2\beta\sqrt{\alpha}}{3}}$, and thus  we get that 
$$\max_{x(t)} x(t)\left(\frac{3}{\beta}\sqrt{\alpha n_{\textsf{alg}}(t)} -\alpha x(t)\right) = 
 \frac{n_{\textsf{alg}}(t)}{\left( \frac{2\beta}{3}\right)^2}.$$
 To upper bound the second term, we maximize $\alpha \sum_{t\in T_i}  \left(x(t-1)^2 - 2 x(t)x(t-1))\right)$  subject to the work neutrality constraint, $\sum_{t\in T_i}x(t) = \sfA_i$. Since it is a convex program, using the KKT conditions, the optimal $x(t)-x(t-1) = \frac{\lambda^\star}{2},$ where $\lambda^\star = \frac{4\sfA_i}{(t_{i+1}-t_i)(t_{i+1}-t_i-1)}$. Thus, using a bit of algebra, we get that under the work neutrality constraint,
 $$\alpha \sum_{t}  \left(x(t-1)^2 - 2 x(t)x(t-1))\right) \le \frac{4 \alpha \sfA_i^2}{(t_{i+1}-t_i)^2}.$$
 For $\textsf{alg}$, either the length of a cycle $t_{i+1}-t_i = \Omega(\frac{\sqrt{\alpha \sfA_i}}{\beta})$ or $t_{i+1}-t_i = \Omega(\sfA_i)$. In both cases,  
 $$\alpha \sum_{t\in T_i}  \left(x(t-1)^2 - 2 x(t)x(t-1))\right) \le \frac{4 \alpha \sfA_i^2}{(t_{i+1}-t_i)^2} = o(F_\textsf{alg} (C_i)),$$
 where  $F_\textsf{alg} (C_i)$  is the flow time of jobs with $\textsf{alg}$ that arrive during $C_i$. 
  
 From \eqref{eq:dummy}, summing over 
all cycles we get that 
$$\max_{x(t), \ d_j, \ \mathsf{acc}(t)\le D(t)} \sum_{t} x(t)\left(\frac{3}{\beta} (\alpha n_{\textsf{alg}}(t))^{1/2} - \alpha x(t)\right) + \alpha \sum_{t}  \left(x(t-1)^2 - 2 x(t)x(t-1))\right)$$
 \begin{equation}\label{eq:objinterest2}
  \le \sum_t \frac{n_{\textsf{alg}}(t)}{\left( \frac{2\beta}{3}\right)^2} + \sum_{i:C_i} o(F_\textsf{alg} (C_i)) \le \frac{9}{4\beta^2}F_\textsf{alg} + o(F_\textsf{alg}).
\end{equation} 

The other possibility is that instead of making $\sum_{t\in T_i}x(t) = \sfA_i$ for each busy cycle $i$ (i.e. exhausting the full quota permissible while satisfying the work neutrality constraint in each cycle of $\textsf{alg}$), some resource $x(t)$ is moved  to a future cycle of $\textsf{alg}$. 
In this case, consider the earliest busy cycle $C_{i^\star}$ of $\textsf{alg}$ where $\sum_{t=0}^{t_{i^\star+1}}x(t) = \sum_{j\le i^\star} \sfA_j$, and again using the process as above we obtain,  
$$\max_{x(t), \sum_{t=0}^{t_{i^\star+1}}x(t) = \sum_{j\le i^\star} \sfA_j} \sum_{t =0}^{t_{i^\star+1}} x(t)\left(\frac{3}{\beta} (\alpha n_{\textsf{alg}}(t))^{1/2} - \alpha x(t)\right) + \alpha \sum_{t=0} ^{t_{i^\star+1}} \left(x(t-1)^2 - 2 x(t)x(t-1))\right)$$
\begin{equation}\label{eq:objinterest3}
 \le \frac{9}{4\beta^2} \sum_{j=1}^{i^\star}F_\textsf{alg}(C_j) + \sum_{j=1}^{i^\star} o(F_\textsf{alg}(C_j)).
\end{equation} 

Summing over all cycles  we get 
$$\max_{x(t), \ d_j, \ \mathsf{acc}(t)\le D(t)} \sum_{t} x(t)\left(\frac{3}{\beta} (\alpha n_{\textsf{alg}}(t))^{1/2} - \alpha x(t)\right) + \alpha \sum_{t}  \left(x(t-1)^2 - 2 x(t)x(t-1))\right)$$
\begin{equation}\label{eq:objinterest4}
 \le \frac{9}{4\beta^2} F_\textsf{alg} + o(F_\textsf{alg}).
\end{equation}
 
Recalling the choice of $\lambda_j w_j = \Delta_j F_{\textsf{alg}}$, and using the bound that we have derived in \eqref{eq:objinterest2} and \eqref{eq:objinterest4}, ignoring the lower order terms, we get from \eqref{eq:lbdual2} 

\begin{align}\label{eq:lbdual3} \min_{x(t), \ d_j, \ \mathsf{acc}(t)\le D(t)} \cL (\lambda_j, \mu_j=0, \nu_t=0)   & \ge  \sum_j \lambda_j w_j -  \frac{9}{4\beta^2}F_\textsf{alg} = \sum_{j } \Delta_j F_\textsf{alg} -  \frac{9}{4\beta^2}F_\textsf{alg}=F_\textsf{alg}\left(1-\frac{9}{4\beta^2}\right).
\end{align}

From the weak duality, the primal value of \eqref{eq:probquad} is at least as much as the value of any dual feasible solution. Thus, the competitive ratio of 
any online algorithm $\cA$ is at most the ratio of an upper bound on the cost of $\cA$ and a lower bound on the dual feasible solution. Hence combining  \eqref{eq:lbdual3} and Lemma \ref{eq:costalg}, we get that the competitive ratio of $\textsf{alg}$ is at most 

$$\mu_{\textsf{alg}} \le \frac{1+2\beta^2}{\left(\frac{4\beta^2-9}{\beta}\right)}.$$
For $\beta^2=3$, we get $\mu_{\textsf{alg}} \le 7\times 4= 28,$
while optimizing over $\beta$, we get that the optimal $\beta = 2.177$ and the competitive ratio is $19.95$.
This completes the proof of Theorem \ref{thm:main}.

%% file: AppLemmaFlowTime.tex
\section{Proof of Lemma \ref{lem:lambdajtime}}

\begin{lemma}\label{lem:DeltaF} $\Delta_j F_{\textsf{alg}} \le  ({\hat d}_j-a_j),$
where ${\hat d}_j$ is the completion time of job $j$ (following $\textsf{alg}$) assuming no further jobs arrive after job $j$.
\end{lemma}
\begin{proof} Note that with $\textsf{alg}$ the arrival of a new job does not decrease the number of active servers. Moreover, since all job sizes are equal and $\textsf{alg}$ is following multi-server SRPT, job $j$ does not preempt (block) any of the remaining jobs at time $t$ ever, and the increase in the total flow time is just the flow time of job $j$, i.e. ${\hat d}_j-a_j$.
\end{proof}

Using Lemma \ref{lem:DeltaF} together with the definition of $\lambda_j w_j = \Delta_j F_{\textsf{alg}}$, we get the following result.
\begin{lemma}\label{lem:lambdaj}
$\lambda_j \le  \frac{{\hat d}_j-a_j}{w_j}$ assuming that no new job arrives after time $a_j$.
\end{lemma}

\begin{remark} In contrast to ${\hat d}_j$, recall that $d_j$ is the actual completion time of job $j$ following $\textsf{alg}$ when jobs arrive after time $a_j$ as well. 
Since more jobs can arrive after time $a_j$, $d_j\ne {\hat d}_j$. It is possible that $d_j< {\hat d}_j$, for example when more jobs arrive in near future, the number of active servers can be increased and it is possible that job $j$ might finish faster. 
\end{remark}
To proceed further, next, we record an important property of the $\textsf{alg}$, for which we need the following definition.

\begin{definition}\label{defn:monotone} [Monotonicity] Consider two job arrival instances $\cI$ and $\cI'$ that are identical except that there is a single job $j \in \cI'\backslash \cI$. Moreover, assume that no job arrives after job $j$ that arrives at time $a_j$ in either $\cI$ or $\cI'$. Then an algorithm is defined to be {\bf monotone} if $n^{\cI}(t) \le n^{\cI'}(t)$, where $n^{\cI}(t)$ is the number of remaining jobs at time $t$ with job arrival instance $\cI$.
\end{definition}

\begin{lemma}\label{lem:monotone} When $w_j=w, \ \forall \ j$, $\textsf{alg}$ is monotone.
\end{lemma}

\begin{proof}
From Definition \ref{defn:monotone}, for the two job arrival instances $\cI$ and $\cI'$, with $\textsf{alg}$ $n^{\cI}(t) = n^{\cI'}(t)$  for $t < a_j$, where $a_j$ is the arrival time of job $j$ under consideration.

Note that with $\textsf{alg}$, the arrival of new job cannot decrease the number of active servers. If the number of active servers remains the same with both $\cI$ and $\cI'$ throughout, then the claim follows directly. 

Thus, 
assume that the arrival of job $j$ increases the number of active servers $s(t)$ with the $\textsf{alg}$.  
Let job $j$'s arrival make $s(a_j) = s(a_j-1) + 1$. Thus, an additional server is now active compared to if job $j$ would not have arrived. 
Either job $j$ begins service on the newly active server, which means that there is no job that has arrived before job $j$ and  that is waiting at time $a_j$ to being serviced since all jobs are of equal size and multi-server SRPT is being executed. Thus, the arrival of new job has no effect on the completion time of any other outstanding job with the $\textsf{alg}$. 
Thus, given that 
$n^{\cI}(t) = n^{\cI'}(t)$  for $t< a_j$ by definition, we also have  $n^{\cI}(t) \le n^{\cI'}(t)$ for all $t$.

Now consider that case some other job (say $j'$) that is waiting at time $a_j-1$ begins service on the newly active server at time $a_j$. Then, from the choice of $s(t)$ made by $\textsf{alg}$, as soon as any one of the jobs that is under process at time $a_j$ is complete, the number of active servers reduces to $s(a_j-1)$ (or even lower if more than one jobs completes). Hence it is possible that job $j'$ completes faster because of the arrival of job $j$, however, in terms of the number of jobs in the system, $j$ has taken the place of $j'$ and we again get that $n^{\cI}(t) \le n^{\cI'}(t)$ for all $t$.
\end{proof}

Next, we show an important result that bounds the dual variable $\lambda_j$ in terms of the number of remaining jobs $n_{\textsf{alg}}(t)$ at time $t$.

\begin{lemma}\label{lem:lambdajtime} For $\textsf{alg}$, $\lambda_j - \left(\frac{t-a_j}{w_j}\right) \le \frac{3}{\beta} (\alpha n_{\textsf{alg}}(t))^{1/2}$ for all $t$.
\end{lemma}
For proving Lemma \ref{lem:lambdajtime} we do not need to assume that all job sizes are equal, i.e. Lemma \ref{lem:lambdajtime} is valid even if $w_j\ne w_k$ for $j\ne k$.

To prove Lemma \ref{lem:lambdajtime}, we will need the following intermediate result that bounds the flow time of job $j$ having size $w_j$ with $\textsf{alg}$ when $n-1$ other jobs arrive together with the $j^{th}$ job   at time $0$.

\begin{lemma}\label{lem:ubflowtime} Let $n$ jobs be available at time $0$, and the $j^{th}$ job among them be of size $w_j$. When no further job arrives after time $0$, then for the $\textsf{alg}$, job $j$'s flow time $f_j \le 3 w_j \sqrt{\alpha n}/\beta$.
\end{lemma}
It is worth noticing that Lemma \ref{lem:ubflowtime} does not require all jobs to have the same size. 

\begin{proof} Since the $\textsf{alg}$ uses multi-server SRPT for scheduling jobs and the number of active servers only depend on the number of remaining jobs, the worst case for maximizing the flow time of a job with size $w_j$ is when all other $n-1$ jobs are also of size $w_j$. In Proposition \ref{prop:ubcomptime}, we show that for the $\textsf{alg}$, when $n$ jobs of equal size $w$ are available at time $0$ and no further jobs arrive, all jobs are complete by at most time $\frac{3}{\beta} w \sqrt{n}$. This implies that job $j$ is complete by time $\frac{3}{\beta} w_j \sqrt{n}$ and the result follows.
\end{proof}

\begin{proposition}\label{prop:ubcomptime} Let at time $0$, the number of jobs be $n$ each with size $w$. When no further jobs arrive in future, with the $\textsf{alg}$ the time by which all jobs are finished is at most $\frac{3w}{\beta}\sqrt{\alpha n}$ when $\beta \sqrt{n/\alpha} = \min\{\beta \sqrt{n/\alpha}, n\}$ and $w$ otherwise.
\end{proposition}

\begin{proof}
 First consider $\beta=1$. How to extend it for general $\beta$ is detailed at the end.  Also, without loss of generality let $w=1$, since all jobs are of same size, we will simply scale the result by $w$.
We will prove the Proposition by showing the following two claims: i) $\textsf{alg}$ successively uses $L, L-1, L-2, \dots, 1 $ number of active servers, where $L = \lfloor\sqrt{n/\alpha}\rfloor$, and each number of active servers can be used for multiple consecutive time slots, and 
ii) the number of consecutive time slots for which a particular integral number of servers is used is at most $3$.

Note that by the definition of $\textsf{alg}$, the active number of servers chosen at the beginning is $L = \lfloor\sqrt{n/\alpha}\rfloor$.

To prove claim i), let time $t$ be the first time when the number of active servers is $q$ (some integer) with  the algorithm, and let $n(t)$ be the number of jobs at time $t$. 
Thus, we get that $n(t)/\alpha < (q+1)^2$ since otherwise the number of active servers chosen would have been $q+1$. 
Next, let $t'\ge t$ be the earliest time such that number of active servers $q$ becomes infeasible, i.e. with $t'-t=k$, 
$k$ is the smallest integer for which $n(t) - k q < \alpha q^2$. Therefore we get that $n(t) - (k-1) q \ge \alpha q^2$ which implies that 
$n(t) - k q \ge \alpha q^2 - q$. Since $\alpha q^2-q \ge \alpha (q-1)^2$ for $q\ge 1$, we get that $n(t) - k q \ge \alpha (q-1)^2$. Thus, at time $t'$ the number of active servers chosen is $(q-1)$.  
 
 For proving claim ii) again consider time $t$ where  $q$ (some integer) is the active number of servers chosen by the algorithm for the first time. Let $t'\ge t$ be the earliest time after $t$ where the number of active servers chosen by the algorithm is at most $q-1$. We want to show that $t'-t\le 3\alpha$. To prove this, suppose $t'-t>3 \alpha$. Then from time $t$ till $t'$ the number of active servers is $q$. By definition of time $t$, as above, we  get that  
$n(t)/\alpha < (q+1)^2$. Choosing active number of servers $q$ for $3\alpha$ consecutive slots, we get the number of jobs at slot $t+3$ to be $n(t) - 3\alpha q $ which is less than $\alpha q^2$ for any positive integer $q$, given that $n(t)/\alpha < (q+1)^2$. Thus at slot $t+3\alpha$ active number of servers  $q-1$ will be chosen, implying that $t'-t\le 3\alpha$.

Claim i) and ii) together imply that the total number of time slots needed to finish $n$ jobs of size $w$ (which is also the total flow time we are interested in) with $\beta=1$ is at most $3\sqrt{\alpha n} w$.  

To extend the result for any $\beta>1$, note that when $\beta \sqrt{n/\alpha} = \min\{\beta \sqrt{n/\alpha}, n\}$,  the flow time upper bound (derived above) just  gets scaled by $\frac{1}{\beta}$. Let $n =  \min\{\beta \sqrt{n/\alpha}, n\}$. In this case, all $n$ jobs are processed simultaneously and the time by which all jobs will finish with $\textsf{alg}$ is $w$.
\end{proof}

Now we are ready to prove Lemma \ref{lem:lambdajtime}.

\begin{proof} [Proof of Lemma \ref{lem:lambdajtime}:]
From the monotone property of the $\textsf{alg}$ (Lemma \ref{lem:monotone}), it is sufficient to prove this result assuming that no job arrives after time $a_j$, the arrival time of job $j$ with size $w_j$. Also, recall that $\lambda_j w_j = \Delta_j F \le  {\hat d}_j-a_j$. Since no job arrives after job $j$, we also have  $d_j={\hat d}_j$. Moreover, from the fact that the $\textsf{alg}$ follows multi-server SRPT policy and the number of active servers only depend on the number of remaining jobs, the flow time of job $j$, $d_j-a_j$, is maximized if all the remaining jobs at time $a_j$ are of size $w_j$. Thus, we will consider this from here on. 

Let $S$ be the set of jobs remaining at time $a_j$ including $j$. Since job $j$ will finish the last as all the remaining jobs at time $a_j$ are of size $w_j$ and job $j$ has arrived the last,  with abuse of notation we let $|S|=j$. Consider time $t\ge a_j$, where $u-1$ jobs have been completed among $S$. Among the jobs in set $S$, let $d_i$ be the time of departure of the $i^{th}$ job in order. Then by the definition of $u$, we get that $[a_j, d_1) \cup [d_1, d_2) \cup \dots \cup [d_{u-2}, d_{u-1}) \subset [a_j, t]$. 

Thus \begin{equation}\label{eq:dummy2}
t-a_j \ge (d_1-a_j) + (d_2-d_1) +\dots + (d_{u-1}-d_{u-2}).
\end{equation}

Since $d_j={\hat d}_j$, from Lemma \ref{lem:lambdaj}, we have
$$\lambda_j - \frac{t-a_j}{w_j} \le \frac{ d_j-a_j}{w_j} - \frac{t-a_j}{w_j}.$$

Using  \eqref{eq:dummy2}, we get that 

$$\lambda_j - \frac{t-a_j}{w_j} \le \frac{(d_j-d_{j-1}) + (d_{j-1}-d_{j-2}) +\dots + (d_{u}-d_{u-1})}{w_j}.$$
By definition of $u$, the number of jobs remaining at time $t$ is $n(t) = j-u+1$, and $(d_j-d_{j-1}) + (d_{j-1}-d_{j-2}) +\dots + (d_{u}-d_{u-1})$ is at most the time needed to finish 
$n_{\textsf{alg}}(t) = j-u+1$ jobs of size at most $w_j$ each, with the $\textsf{alg}$. This time can be largest if the number of servers used by the $\textsf{alg}$ is $0$ at time $t$. Thus, applying Lemma \ref{lem:ubflowtime} where the initial number of active servers  is $0$, we get that $(d_j-d_{j-1}) + (d_{j-1}-d_{j-2}) +\dots + (d_{u}-d_{u-1}) \le \frac{3}{\beta} w_j \sqrt{\alpha (j-u+1)} = \frac{3}{\beta} w_j \sqrt{\alpha n_{\textsf{alg}}(t)}$ as required.
\end{proof}